\documentstyle[12pt]{article}
\newcommand{\beq}{\begin{equation}}
\newcommand{\eeq}{\end{equation}}
\newcommand{\bea}{\begin{eqnarray}}
\newcommand{\eea}{\end{eqnarray}}
\newcommand{\beas}{\begin{eqnarray*}}
\newcommand{\eeas}{\end{eqnarray*}}
\newcommand{\nn}{\nonumber}

\newcommand{\limit}{\rightarrow}

\newcommand{\tr}{{\rm tr}}

\newcommand{\bra}{\langle}
\newcommand{\ket}{\rangle}
\renewcommand{\theequation}{\arabic{section}.\arabic{equation}}
\begin{document}
\topmargin 0pt
\oddsidemargin 5mm
\headheight 0pt
\topskip 0mm
%%%%%%%%%%%%%%%%%%%%%%%% Title Page %%%%%%%%%%%%%%%%%%%%%%%%%%%%%%%
\begin{titlepage}
\begin{flushright}
December 1999 \\
KEK-TH-667 \\
hep-th/9912254
\end{flushright}
\vspace{1cm}
\begin{Large}
\begin{center}
{\sc Witten's Open String Field Theory} \\
{\sc in Constant $B$-Field Background} \\
\end{center}
\end{Large}
\vspace{5mm}
\begin{center}
\begin{large}
{\sc Fumihiko Sugino}\footnote{E-mail address: {\tt sugino@post.kek.jp}}\\
\end{large}
\vspace{4mm}
{\it Institute of Particle and Nuclear Studies,}\\
{\it High Energy Accelerator Research Organization (KEK),}\\
{\it Tsukuba, Ibaraki 305-0801, Japan}\\
\vspace{3cm}
\begin{large} 
Abstract 
\end{large}
\end{center}

  In this paper we consider Witten's bosonic open string field theory 
in the presence of a constant background of the second-rank antisymmetric tensor 
field $B_{ij}$. We extend the operator formulation of Gross and Jevicki 
in this situation and construct the overlap vertices explicitly. 
As a result we find a noncommutative structure of the Moyal type 
only in the zero-mode sector, which is consistent with 
the result of the correlation functions among vertex operators 
in the world sheet formulation. 
Furthermore we find out a certain unitary transformation of the string field 
which absorbs the Moyal type noncommutative structure. 
It can be regarded as a microscopic origin of the transformation 
between the gauge fields in commutative and noncommutative gauge theories discussed by 
Seiberg and Witten.

\end{titlepage}

%%%%%%%%%%%%%%%%%%%%%%%% Section 1: Introduction %%%%%%%%%%%%%%%%%%%%%%%%

\section{Introduction}

  Noncommutative nature of space-time has often appeared in nonperturbative aspects 
of string theory. It has been used in a formulation of 
interacting open string field theory by Witten \cite{Witten,WittenS}. 
He has written a classical action of open string field theory in terms of 
noncommutative geometry, where the noncommutativity appears in a product of string fields. 
Later, four years ago, 
the Dirichlet branes (D-branes) have been recognized as solitonic objects 
in superstring theory \cite{Polchinski}. Further it has been found that the low energy 
behavior of the D-branes are well described 
by supersymmetric Yang-Mills theory (SYM) \cite{Witten1}. 
In the situation of some D-branes coinciding, 
the space-time coordinates are promoted to matrices 
which appear as the fields in SYM. 
Then the size of the matrices corresponds to the number of the D-branes, so 
noncommutativity of the matrices is related to the noncommutative nature of space-time. 
On the other hand, it is known that 
when considering open strings in the presence of a constant background of the 
second-rank antisymmetric tensor field $B_{ij}$, 
the end points of the open strings become noncommutative \cite{ACNY,ChuHo,AAS}. 
It means a noncommutativity of the world volume coordinates of the D-branes, 
which occurs even for a single D-brane differently from the situation mentioned above. 
Although these two types of noncommutative natures have distinct origins, 
as was explicitly pointed out in refs. \cite{Ishibashi1,Kato-Kuroki}, 
they are related to each other in the interesting way: 
``Infinitely many coincident D$p$-branes can be described by a single D($p+2$)-brane 
under a constant $B$-field background." 

 Recently, Seiberg and Witten \cite{SW} 
discussed that the low energy behavior of the D-branes 
in the constant $B$-field background can be described either by commutative or 
noncommutative Yang-Mills theories\footnote{The terminology ``commutative Yang-Mills 
theory" or ``noncommutative Yang-Mills theory" means that the geometry of the 
space-time where the Yang-Mills theory is defined on is commutative or noncommutative 
respectively.}, which corresponds to the difference of the regularization scheme --- 
the Pauli-Villars regularization or the point-splitting regularization --- 
adopted in the world sheet approach to open string theory.  
Also they argued the relation between the gauge field in the commutative theory and 
that in the noncommutative theory, and obtained its explicit form 
in some simple cases\footnote{For related works, 
see refs. \cite{Okawa,Ishibashi2,Okuyama}.}. 

In this paper we consider Witten's open string field theory in a constant background of 
the $B$-field. Then we should have a new noncommutativity originating form 
the $B$-field background in addition to the noncommutativity which already has appeared 
in the product of the string fields. 
We consider the open bosonic string field theory instead of supersymmetric one. 
Because the noncommutative 
structure mentioned above appears independently of supersymmetry, 
we do not consider the fermions for the sake of simplicity.  
We explicitly construct this open string theory by extending the operator formulation, 
which has been given by Gross and Jevicki \cite{GJ1,GJ2,GJ3} 
in the background of the flat space-time, to the case of the presence of 
the constant $B$-field 
background. As a result we obtain a product between the string fields added to 
the ordinary product by the Moyal 
product acting only on the zero-modes of the open strings, 
which reproduces the result of the correlation functions among vertex operators 
in the world sheet formulation \cite{SW}. 
It is noted that the part of the product concerning the nonzero-modes is not affected 
by the presence of the $B$-field background. 
The string field theory with the modified product gives noncommutative Yang-Mills 
theory in the low energy limit. 
In the string field theory perspective, 
there should exist a certain (hopefully simpler than the Yang-Mills case) 
transformation connecting the 
string field in the string field theory we have constructed and that in a string field 
theory leading to commutative Yang-Mills theory. 
Here we find that the Moyal type noncommutative structure can be absorbed 
by a unitary rotation of the string field. As a result of the rotation, 
the string field theory is transformed to a string field theory with the ordinary 
noncommutative product where the BRST charge in the kinetic term is unaffected. 
This fact suggests that the universality of the interaction vertices 
in various string field theories, which has been discussed 
in refs. \cite{Kugo-Zwiebach,Yoneya,HLRS}, 
holds also between open string field theories with 
different boundary conditions caused by turning on the $B$-field. 
The string field theory obtained by the transformation 
can be considered to give a microscopic origin of the commutative Yang-Mills theory.

  This paper is organized as follows. 
In section 2, we briefly review Witten's bosonic open string field theory 
and its explicit 
construction based on the operator formulation elaborated by Gross and Jevicki 
in the case of the Neumann boundary condition. 
We explain open strings in the presence of a constant $B$-field background 
and obtain the mode-expanded form of the open string in section 3. 
As a result, 
we see that in this situation the end points of the open string become noncommutative.   
It has already been discussed by many authors \cite{ACNY,ChuHo,AAS,SW}. But in order to 
make this paper more self-contained and to prepare for construction of 
Witten's string field theory in the constant $B$-field background, we give a 
somewhat detailed argument. 
In section 4, based on the results in the previous sections, Witten's open string field 
theory under the constant $B$-field background is constructed explicitly by making use of 
the operator formulation. Here we obtain 
the structure of the Moyal product in addition to the ordinary noncommutativity 
which has already existed in the background of no $B$-field. 
In section 5, we find a certain unitary transformation of the string field 
which removes the Moyal type noncommutativity from the string field theory 
constructed in the previous section. This transformation can be essentially regarded 
as a microscopic analog 
of the map between the gauge fields in 
commutative and noncommutative Yang-Mills theories 
discussed in ref. \cite{SW}. 
In section 6, we also discuss about the background independence 
between open string field theories with different boundary conditions caused by 
the presence of the constant $B$-field. 
Until now we have not put the Chan-Paton factors to the end points of 
open strings, which corresponds to the case of the $U(1)$ gauge group 
in the Yang-Mills theories. 
However, in string field theory, 
it is quite a straightforward generalization 
to introduce the Chan-Paton factors as is pointed out 
in refs. \cite{Dearnaley1,Dearnaley2}, 
so we can immediately have the results also in the case 
corresponding to the nonabelian gauge group. 
It forms a contrast to the situation in Yang-Mills theory, 
where in general 
it is difficult to obtain the concrete form of the map connecting the gauge fields of 
the commutative and noncommutative Yang-Mills theories 
except some simple cases\footnote{For example, 
the case of the $U(1)$ gauge fields giving constant field strengths.}. 
Finally, section 7 is devoted to summary and discussion. 
In appendix A we present the result of the explicit forms of the overlap vertices 
discussed in section 2.

%%%%%%%%%%%%%%%%%%%%%%%% Section 2: Brief Review %%%%%%%%%%%%%%%%%%%%%%%%

\section{Brief Review of Witten's Open String Field Theory}

In this section, we review some basic properties of Witten's 
bosonic open string field theory \cite{Witten} and its explicit 
construction based on a Fock space representation of string field 
functional and $\delta$-function overlap vertices \cite{GJ1,GJ2,CST}. 

\subsection{Witten's Open String Field Theory}

Witten introduced a beautiful formulation of open string field theory 
in terms of a noncommutative extension of differential geometry, 
where string fields, the BRST operator $Q$ and the integration over 
the string configurations $\int$ in string field theory are analogs of 
differetial forms, the exterior derivative $d$ and the integration over 
the manifold $\int_M$ in the differential geometry, respectively. 
The ghost number assigned to the string field corresponds to 
the degree of the differential form. 
Also the (noncommutative) product between the string fields $*$ is 
interpreted as an 
analog of the wedge product $\wedge$ between the differential forms. 

The axioms obeyed by the system of $\int$, $*$ and $Q$ are 
\bea 
 & & \int Q A=0, \nn \\
 & & Q(A*B) = (QA)*B+(-1)^{n_A}A*(QB), \nn \\
 & & (A*B)*C = A*(B*C), \nn \\
 & & \int A*B = (-1)^{n_An_B}\int B*A, 
\eea
where $A$, $B$ and $C$ are arbitrary string fields, 
whose ghost number is half-integer valued: 
The ghost number of $A$ is defined by the integer $n_A$ as 
$n_A+\frac12.$ 

 Then Witten discussed the following string field theory action 
\beq
S=\frac{1}{G_s}\int \left(\frac12 \psi *Q\psi+\frac13 \psi*\psi*\psi\right), 
\label{SFTaction1}
\eeq
where $G_s$ is the open string coupling constant and $\psi$ is 
the string field with the ghost number $-\frac12.$ 
The action is invariant under the gauge transformation 
\beq
\delta\psi = Q\Lambda+\psi*\Lambda-\Lambda*\psi, 
\eeq
with the gauge parameter $\Lambda$ of the ghost number $-\frac32.$

\subsection{Operator Formulation of String Field Theory}

The objects defined above can be explicitly constructed 
by using the operator formulation, where the string field is represented 
as a Fock space, and the integration $\int$ as an inner product on the 
Fock space. It was considered by Gross and Jevicki \cite{GJ1,GJ2} 
in the case of the Neumann boundary condition. 
We will heavily use the notation of the papers \cite{GJ1,GJ2}. 
In the operator formulation, the action (\ref{SFTaction1}) is described as 
\beq
S=\frac{1}{G_s}\left(\frac12 \mbox{ }_{12}\!\bra V_2||\psi\ket_1 Q|\psi\ket_2
+\frac13 \mbox{ }_{123}\!\bra V_3||\psi\ket_1 |\psi\ket_2 |\psi\ket_3\right), 
\label{SFTaction2}
\eeq
where the structure of the product $*$ in the kinetic and potential terms 
is encoded to that of the overlap vertices 
$\bra V_2|$ and $\bra V_3|$ respectively\footnote{Subscripts put to vectors 
in the Fock space label the strings concerning the vertices.}. 

As a preparation for giving the explicit form of the overlaps, 
let us consider open strings in 26-dimensional space-time with the constant metric 
$G_{ij}$ in the Neumann boundary condition. 
The world sheet action is given by 
\beq
S_{WS}=\frac{1}{4\pi\alpha'}\int d\tau \int_0^{\pi}d\sigma G_{ij}
(\partial_{\tau}X^i\partial_{\tau}X^j-\partial_{\sigma}X^i\partial_{\sigma}X^j)
+S_{gh}, 
\label{wsactionG}
\eeq
where $S_{gh}$ is the action of the $bc$-ghosts: 
\beq
S_{gh}= \frac{i}{2\pi}\int d\tau\int_0^{\pi}d\sigma
[c_+(\partial_{\tau}-\partial_{\sigma})b_{+}
+c_{-}(\partial_{\tau}+\partial_{\sigma})b_{-}].
\label{ghostaction}
\eeq
Under the Neumann boundary condition, the string coordinates have the standard mode 
expansions: 
\beq
X^j(\tau,\sigma)=x^j+2\alpha'\tau p^j+
i\sqrt{2\alpha'}\sum_{n\neq 0}\frac1n\alpha_n^je^{-in\tau}\cos (n\sigma), 
\label{modeexpansionN}
\eeq
also the mode expansions of the ghosts are given by 
\bea
c_{\pm}(\tau,\sigma) & = & \sum_{n\in{\bf Z}}c_ne^{-in(\tau\pm\sigma)}
\equiv c(\tau,\sigma)\pm i\pi_b(\tau,\sigma), \nn \\
b_{\pm}(\tau,\sigma) & = & \sum_{n\in{\bf Z}}b_ne^{-in(\tau\pm\sigma)}
\equiv \pi_c(\tau,\sigma)\mp ib(\tau,\sigma). 
\eea
As a result of the quantization, the modes obey the commutation relatons: 
\bea
 & & [x^i,p^j]=iG^{ij}, \hspace{1cm} [\alpha_n^i, \alpha_m^j]=nG^{ij}\delta_{n+m,0}, \\
 & & \{b_n, c_m\}=\delta_{n+m,0}, 
\eea
the otherwises vanish. 

The overlap 
$$
|V_N\ket =|V_N\ket^X|V_N\ket^{gh} \hspace{1cm}(N=1,2,\cdots) 
$$
is the state satisfying 
the continuity conditions for the string coordinates and the ghosts 
at the $N$-string vertex of the string field theory. The superscripts $X$ and $gh$ show 
the contribution of the sectors of the coordinates and the ghosts respectively. 
The continuity conditions for the coordinates are 
\bea
 & & (X^{(r)j}(\sigma)-X^{(r-1)j}(\pi-\sigma))|V_N\ket^X =0, \nn \\
 & & (P^{(r)}_i(\sigma)+P^{(r-1)}_i(\pi-\sigma))|V_N\ket^X =0, 
\label{GJ1condition}
\eea
for $0\le \sigma \le\frac{\pi}{2}$ and $r=1,\cdots,N$. 
Here $P_i(\sigma)$ is the momentum conjugate to the coordinate $X^j(\sigma)$ at $\tau=0$, 
and the superscript $(r)$ labels the string $(r)$ meeting at the 
vertex\footnote{We often 
use the expression $|V_N\ket_{12\cdots N}$ for the overlap when specifying the strings 
concerning the vertex.}. 
In the above formulas, we regard $r=0$ as $r=N$ because of the cyclic property of 
the vertex. 
For the ghost sector, we impose the following conditions on the variables 
$c(\sigma)$, $b(\sigma)$ and their conjugate momenta $\pi_c(\sigma)$, $\pi_b(\sigma)$: 
\bea
 & & (\pi_c^{(r)}(\sigma)-\pi_c^{(r-1)}(\pi-\sigma))|V_N\ket^{gh} =0, \nn \\
 & & (b^{(r)}(\sigma)-b^{(r-1)}(\pi-\sigma))|V_N\ket^{gh} =0, \nn \\
 & & (c^{(r)}(\sigma)+c^{(r-1)}(\pi-\sigma))|V_N\ket^{gh} =0, \nn \\
 & & (\pi_b^{(r)}(\sigma)+\pi_b^{(r-1)}(\pi-\sigma))|V_N\ket^{gh} =0, 
\label{GJ2condition}
\eea
for $0\le \sigma \le\frac{\pi}{2}$ and $r=1,\cdots,N$.

We present the explicit form of $|V_N\ket$ for $N=1,2,3,4$ in appendix A.

%%%%%%%%%%%%%%%%%% Section 3: Open Strings in B Background %%%%%%%%%%%%%
\section{Open Strings in Constant $B$-Field Background}
\setcounter{equation}{0}
  We consider a constant background of the second-rank antisymmetric tensor 
field $B_{ij}$ in addition to the constant metric $g_{ij}$ where open strings 
propagate. Then the boundary conditon at the end points of the open strings 
changes from the Neumann type, 
and thus the open string has a different mode expansion from 
the Neumann case (\ref{modeexpansionN}). 
As a result, the end point is to be noncommutative, 
in the picture of the D-branes 
which implies noncommutativity of the world volume coordinates on the D-branes. 
Here we derive the mode-expanded form of the open string coordinates as a 
preparation for a calculation of the overlap vertices in the next section. 

We start with the world sheet action 
\bea
S_{WS}^B  & =  & \frac{1}{4\pi\alpha'}\int d\tau \int_0^{\pi}d\sigma [
g_{ij}(\partial_{\tau}X^i\partial_{\tau}X^j-\partial_{\sigma}X^i\partial_{\sigma}X^j)
-2\pi\alpha'B_{ij}(\partial_{\tau}X^i\partial_{\sigma}X^j
-\partial_{\sigma}X^i\partial_{\tau}X^j)] \nn \\
 &  & +S_{gh}. 
\label{wsactionB}
\eea
Because the term proportional to $B_{ij}$ can be written 
as a total derivative term, it does not affect the equation of motion but does 
the boundary condition, which requires 
\beq
g_{ij}\partial_{\sigma}X^j-2\pi\alpha'B_{ij}\partial_{\tau}X^j=0 
\label{boundarycondition1}
\eeq
on $\sigma=0,\pi$. This can be rewritten to the convenient form 
\beq
E_{ij}\partial_-X^j=(E^T)_{ij}\partial_+X^j, 
\label{boundarycondition2}
\eeq
where $E_{ij}\equiv g_{ij}+2\pi\alpha'B_{ij}$, and $\partial_{\pm}$ are derivatives 
with respect to the light cone variables $\sigma^{\pm}=\tau\pm\sigma$. 
We can easily see that $X^j(\tau,\sigma)$ satisfying 
the boundary condition (\ref{boundarycondition2}) has the following mode expansion: 
\bea
X^j(\tau,\sigma) & = & \tilde{x}^j+\alpha'\left[(E^{-1})^{jk}g_{kl}p^l\sigma^-
+(E^{-1T})^{jk}g_{kl}p^l\sigma^+\right] \nn \\
 & & +i\sqrt{\frac{\alpha'}{2}}\sum_{n\neq 0}\frac1n
\left[(E^{-1})^{jk}g_{kl}\alpha^l_ne^{-in\sigma^{-}}+
(E^{-1T})^{jk}g_{kl}\alpha^l_ne^{-in\sigma^{+}}\right]. 
\label{modeexpansion1}
\eea
We will obtain the commutators between the modes from the propagator of the 
open strings, which gives another derivation different from the method by 
Chu and Ho \cite{ChuHo} based on the quantization via the Dirac bracket.
When performing the Wick rotation $\tau \limit -i\tau$ and mapping the world sheet 
to the upper half plane $z=e^{\tau+i\sigma},\; \bar{z}=e^{\tau-i\sigma}$ 
($0\le \sigma \le\pi$), the boundary condition (\ref{boundarycondition2}) becomes 
\beq
E_{ij}\partial_{\bar{z}}X^j=(E^T)_{ij}\partial_zX^j, 
\label{boundarycondition3}
\eeq
which is imposed on the real axis $z=\bar{z}$. 
The propagator $\bra X^i(z,\bar{z})X^j(z',\bar{z'})\ket$ satisfying the 
boundary condition (\ref{boundarycondition3})is determined as 
\bea
 \bra X^i(z,\bar{z})X^j(z',\bar{z'})\ket & = & 
-\alpha'\left[g^{ij}\ln|z-z'|-g^{ij}\ln|z-\bar{z'}|\frac{}{}\right. \nn \\
  &  & \left.
+G^{ij}\ln|z-\bar{z'}|^2
+\frac{1}{2\pi\alpha'}\theta^{ij}\ln\frac{z-\bar{z'}}{\bar{z}-z'}+D^{ij}\right], 
\eea
where $G^{ij}$ and $\theta^{ij}$ are given by 
\bea
G^{ij} & = & \frac12(E^{T-1}+E^{-1})^{ij}=(E^{T-1}gE^{-1})^{ij}=
(E^{-1}gE^{T-1})^{ij}, 
\label{G} \\
\theta^{ij} & = & 2\pi\alpha'\cdot\frac12(E^{T-1}-E^{-1})^{ij} \nn \\
 & = & 
(2\pi\alpha')^2(E^{T-1}BE^{-1})^{ij}=-(2\pi\alpha')^2(E^{-1}BE^{T-1})^{ij}. 
\label{Gtheta}
\eea
Also the constant $D^{ij}$ remains unknown from the boundary condition alone. 
However it is an irrelevant parameter, so we can fix an appropriate value.
The mode-expanded form (\ref{modeexpansion1}) is mapped to 
\bea
X^j(z,\bar{z}) & = & \tilde{x}^j-i\alpha'[(E^{-1})^{jk}p_k\ln\bar{z}
+(E^{-1T})^{jk}p_k\ln z] \nn \\
 & & +i\sqrt{\frac{\alpha'}{2}}\sum_{n\neq 0} \frac 1n \left[(E^{-1})^{jk}
\alpha_{n,k}\bar{z}^{-n}+(E^{-1T})^{jk}\alpha_{n,k}z^{-n}\right]. 
\eea
Note that the indices of $p^l$ and $\alpha_n^l$ were lowered by the metric $g_{ij}$ 
not $G_{ij}$. Recall the definition of the propagator 
\beq
\bra X^i(z,\bar{z})X^j(z',\bar{z'})\ket\equiv 
R(X^i(z,\bar{z})X^j(z',\bar{z'}))-N(X^i(z,\bar{z})X^j(z',\bar{z'})), 
\label{defpropagator}
\eeq
where $R$ and $N$ stand for the radial ordering and the normal ordering respectively. 
We take a prescription for the normal ordering 
which pushes $p_i$ to the right and $\tilde{x}_j$ to the left 
with respect to the zero-modes $p_i$ and $\tilde{x}_j$. It corresponds to considering 
the vacuum satisfying 
\beq
p_j|0\ket=\alpha_{n,j}|0\ket=0 \hspace{1cm}(n>0), \hspace{2cm}\bra0|\alpha_{n,j}=0
\hspace{1cm}(n<0), 
\label{vacuumcondition}
\eeq
which is the standard prescription for calculating 
the propagator of the massless scalar field 
in two-dimensional conformal field theory from the operator formalism. 
Making use of eqs. (\ref{defpropagator}), (\ref{vacuumcondition}) and techniques of the 
contour integration, it is easy to obtain the commutators
\beq
[\alpha_{n,i},\alpha_{m,j}]=n\delta_{n+m,0}G_{ij},\hspace{1cm} 
[\tilde{x}^i,p_j]=i\delta ^i_j, 
\eeq
where the first equation holds for all integers with 
$\alpha_{0,i}\equiv \sqrt{2\alpha'}p_i$. The constant $D^{ij}$ is written as 
$\alpha'D^{ij}=-\bra0|\tilde{x}^i\tilde{x}^j|0\ket$, which corresponds to define
the normal ordering with respect to $\tilde{x}^i$'s as 
$$:\tilde{x}^i\tilde{x}^j:=\tilde{x}^i\tilde{x}^j+\alpha'D^{ij}. $$
Let us fix $D^{ij}$ as $\alpha'D^{ij}=-\frac{i}{2}\theta^{ij}$, which is the convention 
taken in the paper \cite{SW}. Then the coordinates $\tilde{x}^i$ 
become noncommutative: 
$$
[\tilde{x}^i,\tilde{x}^j]=i\theta^{ij}, 
$$
but the center of mass coordinates $x^i\equiv\tilde{x}^i+\frac12\theta^{ij}p_j$ 
can be seen to commute each other. 

Now we have the mode-expanded form of the string coordinates 
and the commutation relations between the modes, which are 
\bea
X^j(\tau,\sigma) & = & x^j+
2\alpha'\left(G^{jk}\tau+
\frac{1}{2\pi\alpha'}\theta^{jk}(\sigma-\frac{\pi}{2})\right)p_k \nn \\
 & & +i\sqrt{2\alpha'}\sum_{n\neq0}\frac1ne^{-in\tau}\left[G^{jk}\cos(n\sigma)
-i\frac{1}{2\pi\alpha'}\theta^{jk}\sin(n\sigma)\right]\alpha_{n,k},
\eea
\beq
[\alpha_{n,i},\alpha_{m,j}]=n\delta_{n+m,0}G_{ij},\hspace{1cm} 
[x^i,p_j]=i\delta ^i_j, 
\eeq
with all the other commutators vanishing. 

Also, due to the formula 
\beq
\sum_{n=1}^{\infty}\frac2n\sin(n(\sigma+\sigma'))=\left\{
\begin{array}{ll} \pi-\sigma-\sigma' & \hspace{1cm}(\sigma+\sigma'\neq 0,2\pi) \\
                            0 & \hspace{1cm} (\sigma+\sigma'= 0,2\pi), 
\end{array}\right. 
\eeq
we can see by a direct calculation that the end points of the string 
become noncommutative 
\beq
[X^i(\tau,\sigma),X^j(\tau,\sigma')]= \left\{\begin{array}{ll}
i\theta^{ij} & \hspace{1cm}(\sigma=\sigma'=0) \\
-i\theta^{ij} & \hspace{1cm}(\sigma=\sigma'=\pi) \\
0  &  \hspace{1cm}({\rm otherwises}). \end{array}\right.
\eeq 
On the other hand, 
it is noted that the conjugate momenta have the mode expansion identical with that in the 
Neumann case: 
\bea
P_i(\tau,\sigma) & = & \frac{1}{2\pi\alpha'}
(g_{ij}\partial_{\tau}-2\pi\alpha'B_{ij}\partial_{\sigma})X^j(\tau,\sigma) \nn \\
 & = & \frac{1}{\pi}p_i+\frac{1}{\pi\sqrt{2\alpha'}}\sum_{n\neq 0}e^{-in\tau}
\cos(n\sigma)\alpha_{n,i}. 
\eea
 
At the end of this section, 
we remark that the relations (\ref{G}) and (\ref{Gtheta}) are 
in the same form as a T-duality transformation, 
although the correspondence is a formal sense, 
because we are not considering any compactification of space-time. 
The generalized T-duality transformation, namely $O(D,D)$-transformation, is defined by 
\beq
E'=(aE+b)(cE+d)^{-1}
\label{ODD}
\eeq
with $a$, $b$, $c$ and $d$ being $D\times D$ real matrices. ($D$ is 
the dimension of space-time.) The matrix 
\beq
h=\left(\begin{array}{cc} a & b \\ c & d \end{array}\right)
\eeq
is $O(D,D)$ matrix, which satisfies 
\beq
h^TJh=J
\eeq
where 
$$
J=\left(\begin{array}{cc} 0 & {\bf 1}_D \\ {\bf 1}_D & 0 \end{array}\right). 
$$
The relations (\ref{G}) and (\ref{Gtheta}) correspond to the case of 
the inversion $a=d=0$, $b=c={\bf 1}_D$.

%%%%%%%%%%% Section 4: Construction of Overlap Vertices %%%%%%%%%%%%%%%%%%%

\section{Construction of Overlap Vertices}
\setcounter{equation}{0}
  Here we construct Witten's open string theory in the constant $B$-field background 
by obtaining the explicit formulas of the overlap vertices. 
As is understood from the fact that the action of the ghosts (\ref{ghostaction}) 
contains no background fields, 
the ghost sector is not affected by turning on the $B$-field background. 
Thus we may consider the coordinate sector only. 
First, let us see the mode-expanded forms 
of the coordinates and the momenta at $\tau=0$ 
\bea
X^j(\sigma) & = & G^{jk}y_k+\frac{1}{\pi}\theta^{jk}(\sigma-\frac{\pi}{2})p_k \nn \\
 & & +2\sqrt{\alpha'}\sum_{n=1}^{\infty}\left[G^{jk}\cos(n\sigma)x_{n,k}
+\frac{1}{2\pi\alpha'}\theta^{jk}\sin(n\sigma)\frac1np_{n,k}\right], \\
P_i(\sigma) & = & \frac{1}{\pi}p_i+\frac{1}{\pi\sqrt{\alpha'}}\sum_{n=1}^{\infty}\cos(n\sigma)p_{n,i}, 
\eea
where $x^j=G^{jk}y_k$, the coordinates and the momenta for the oscillator modes are 
\bea
x_{n,k} & = & \frac{i}{2}\sqrt{\frac2n}(a_{n,k}-a_{n,k}^{\dagger})=
\frac{i}{\sqrt2n}(\alpha_{n,k}-\alpha_{-n,k}),  \nn \\
p_{n,k} & = & \sqrt{\frac{n}{2}}(a_{n,k}+a_{n,k}^{\dagger})=
\frac{1}{\sqrt2}(\alpha_{n,k}+\alpha_{-n,k}).
\eea
The nonvanishing commutators are given by 
\beq
[x_{n,k}, p_{m,l}]=iG_{kl}\delta_{n,m}, \hspace{1cm} [y_k,p_l]=iG_{kl}. 
\label{commutatorsB}
\eeq
We should note that the metric appearing in eqs. (\ref{commutatorsB}) is $G_{ij}$, 
instead of $g_{ij}$. So it can be seen that 
if we employ the variables with the lowered space-time indices 
$y_k$, $p_k$, $x_{n,k}$ and $p_{n,k}$, 
the metric used in the expression of the overlaps is $G^{ij}$ not $g^{ij}$. 

  The continuity condition (\ref{GJ1condition}) is universal for any background, 
and the mode expansion of the momenta $P_i(\sigma)$'s is of the same form as in the 
Neumann case, thus the continuity conditions for the momenta in terms of the modes $p_{n,i}$ 
are identical with those in the Neumann case. 
Also, since $p_{n,i}$'s mutually commute, 
it is natural to find a solution of the continuity condition, 
assuming the following form for the overlap vertices: 
\beq
|\hat{V}_N\ket^X_{1\cdots N}= \exp\left[\frac{i}{4\pi\alpha}\theta^{ij}\sum_{r,s=1}^{N}
p^{(r)}_{n,i}Z^{rs}_{nm}p^{(s)}_{m,j}\right]|V_N\ket^X_{1\cdots N}, 
\label{formofoverlap}
\eeq
where $|\hat{V}_N\ket^X_{1\cdots N}$ and $|V_N\ket^X_{1\cdots N}$ 
are the overlaps in the background corresponding to the world sheet actions 
(\ref{wsactionB}) and (\ref{wsactionG}) respectively, 
the explicit form of the latter is given in appendix A. 
Clearly the expression (\ref{formofoverlap}) satisfies the continuity conditions for 
the modes of the momenta, and the coefficients $Z^{rs}_{nm}$'s are determined so that 
the continuity conditions for the coordinates are satisfied.

\subsection{$|\hat{I}\ket^X\equiv|\hat{V}_1\ket^X$}

  For the $N=1$ case, we consider the identity overlap 
$|\hat{I}\ket^X\equiv|\hat{V}_1\ket^X$. 
The continuity conditions for the momenta require that 
\beq
P_i(\sigma)+P_i(\pi-\sigma) = 
\frac{2}{\pi}p_i+\frac{2}{\pi\sqrt{\alpha'}}\sum_{n=2,4,6,\cdots}\cos(n\sigma)p_{n,i}
\eeq
should vanish for $0\leq\sigma\leq\frac{\pi}{2}$, namely, 
\beq
p_i=0, \hspace{1.5cm}p_{n,i}=0 \hspace{1cm}(n=2,4,6,\cdots), 
\label{conditionI1}
\eeq
which is satisfied by the overlap in the Neumann case $|I\ket$. 
In addition, the conditions for the coordinates are that
\bea
X^j(\sigma)-X^j(\pi-\sigma) & = & \frac{2}{\pi}\theta^{jk}(\sigma-\frac{\pi}{2})p_k \nn \\
 & & +4\sqrt{\alpha'}\sum_{n=1,3,5,\cdots}G^{jk}\cos(n\sigma)x_{n,k} \nn \\
 & & +4\sqrt{\alpha'}\sum_{n=2,4,6,\cdots}\frac{1}{2\pi\alpha'}\theta^{jk}\sin(n\sigma)
\frac1np_{n,k}
\label{conditionI2}
\eea
should vanish for $0\leq\sigma\leq\frac{\pi}{2}$. The first and third lines in the r. h. s. can 
be put to zero by using eq. (\ref{conditionI1}). 
So what we have to consider is the remaining 
condition 
$x_{n,k}=0$ for $n=1,3,5,\cdots$, which however is nothing but the continuity condition 
for the coordinates in the Neumann case. It can be understood from the point 
that the second line 
in eq. (\ref{conditionI2}) does not depend on $\theta^{ij}$. 
Thus it turns out that the continuity conditions in the case of the $B$-field turned on 
are satisfied by 
the identity overlap made in the Neumann case (\ref{identityB=0}). The solution is 
\beq
|\hat{I}\ket^X=|I\ket^X=\exp \left[-\frac12G^{ij}\sum_{n=0}^{\infty}(-1)^n a^{\dagger}_{n,i}a^{\dagger}_{n,j}
\right]|0\ket ,
\label{IB}
\eeq
where also the zero modes $y_i$ and $p_i$ are written by using 
the creation and annihilation operators 
$a_{0,i}^{\dagger}$ and $a_{0,i}$ as 
\beq
y_i=\frac{i}{2}\sqrt{2\alpha'}(a_{0,i}-a^{\dagger}_{0,i}), \hspace{1cm}
p_i=\frac{1}{\sqrt{2\alpha'}}(a_{0,i}+a^{\dagger}_{0,i}). 
\eeq

\subsection{$|\hat{V}_2\ket^X_{12}$}

 For the $N=2$ case, we are to do the same argument as in the $N=1$ case. 
The continuity conditions mean that 
\bea
 P^{(1)}_i(\sigma)+ P^{(2)}_i(\pi-\sigma) & = & 
\frac{1}{\pi}(p^{(1)}_i+p^{(2)}_i) 
 +\frac{1}{\pi\sqrt{\alpha'}}\sum_{n=1}^{\infty}\cos(n\sigma)
(p^{(1)}_{n,i}+(-1)^np^{(2)}_{n,i}), \\
 X^{(1)j}(\sigma)-X^{(2)j}(\pi-\sigma) 
 & = & G^{jk}(y^{(1)}_k-y^{(2)}_k)
+\frac{1}{\pi}\theta^{jk}(\sigma-\frac{\pi}{2})(p^{(1)}_k+p^{(2)}_k) \nn \\
 & & \;+2\sqrt{\alpha'}\sum_{n=1}^{\infty}\left[G^{jk}\cos(n\sigma)
(x^{(1)}_{n,k}-(-1)^nx^{(2)}_{n,k})\frac{}{}\right.\nn \\ 
 & & \hspace{2cm}\left.+\frac{1}{2\pi\alpha'}\theta^{jk}\sin(n\sigma)\frac1n
(p^{(1)}_{n,k}+(-1)^np^{(2)}_{n,k})\right] 
\eea
should be zero for $0\leq\sigma\leq\pi$. 
It turns out again that the conditions for the modes are 
identical with those in the Neumann case: 
\bea
 & & p^{(1)}_i+p^{(2)}_i=0, \hspace{1cm} p^{(1)}_{n,i}+(-1)^np^{(2)}_{n,i}=0, \nn \\
 & & y^{(1)}_i-y^{(2)}_i=0, \hspace{1cm} x^{(1)}_{n,i}-(-1)^nx^{(2)}_{n,i}=0, 
\eea
for $n\ge 1$. Thus we have the solution\footnote{For the $N=1$ and 2 cases, 
it turns out that the phase factor in (\ref{formofoverlap}) becomes trivial, 
by using the continuity conditions for the momenta and the antisymmetric property 
of $\theta^{ij}$. We can understand the results (\ref{IB}) and (\ref{V2B}) also from 
this point.} 
\beq
|\hat{V}_2\ket^X_{12}=|V_2\ket^X_{12}=
\exp\left[-G^{ij}\sum_{n=0}^{\infty}(-1)^n a^{(1)\dagger}_{n,i}a^{(2)\dagger}_{n,j}
\right] |0\ket_{12}. 
\label{V2B}
\eeq

\subsection{$|\hat{V}_4\ket^X_{1234}$}

  We find a solution of the continuity conditions (\ref{GJ1condition}) in the $N=4$ case 
assuming the form 
\beq
|\hat{V}_4\ket^X_{1234}=\exp\left[\frac{i}{4\pi\alpha}\theta^{ij}\sum_{r,s=1}^{4}
p^{(r)}_{n,i}Z^{rs}_{nm}p^{(s)}_{m,j}\right]|V_4\ket^X_{1234}. 
\label{V4form}
\eeq
When considering the continuity conditions, it is convenient to employ the $Z_4$-Fourier 
transformed variables: 
\bea
Q_1^j(\sigma) & = & 
\frac12[iX^{(1)j}(\sigma)-X^{(2)j}(\sigma)-iX^{(3)j}(\sigma)+X^{(4)j}(\sigma)] 
\equiv Q^j(\sigma),\nn \\
Q^j_2(\sigma) & = & 
\frac12[-X^{(1)j}(\sigma)+X^{(2)j}(\sigma)-X^{(3)j}(\sigma)+X^{(4)j}(\sigma)], \nn \\
Q_3^j(\sigma) & = & 
\frac12[-iX^{(1)j}(\sigma)-X^{(2)j}(\sigma)+iX^{(3)j}(\sigma)+X^{(4)j}(\sigma)] 
\equiv \bar{Q}^j(\sigma),\nn \\
Q^j_4(\sigma) & = & 
\frac12[X^{(1)j}(\sigma)+X^{(2)j}(\sigma)+X^{(3)j}(\sigma)+X^{(4)j}(\sigma)]. 
\eea
For the momentum variables we also define the $Z_4$-Fourier 
transformed variables $P_{1,i}(\sigma)(\equiv P_i(\sigma))$, 
$P_{2,i}(\sigma)$, $P_{3,i}(\sigma)(\equiv \bar{P}_i(\sigma))$ and 
$P_{4,i}(\sigma)$ by the same combinations of $P^{(r)}_i(\sigma)$'s as the above. 
These variables have the following mode expansions 
\bea
P_{t,i}(\sigma) & = & \frac{1}{\pi\sqrt{2\alpha'}}P_{t,0,i}
+\frac{1}{\pi\sqrt{\alpha'}}\sum_{n=1}^{\infty}\cos(n\sigma)P_{t,n,i}, \nn \\
Q^j_t(\sigma) & = & G^{jk}\sqrt{2\alpha'}Q_{t,0,k}+
\frac{1}{\pi}\theta^{jk}(\sigma-\frac{\pi}{2})\frac{1}{\sqrt{2\alpha'}}P_{t,0,k} \nn \\
 & & +\sqrt{2\alpha'}\sum_{n=1}^{\infty}\left[G^{jk}\cos(n\sigma)Q_{t,n,k}
+\frac{1}{2\pi\alpha'}\theta^{jk}\sin(n\sigma)\frac1nP_{t,n,k}\right], 
\label{modeexpansionZ4}
\eea
where $t=1,2,3,4$. From now on, 
we frequently omit the subscript $t$ for the $t=1$ case, and at the same time 
we employ the expression 
with a bar instead of putting the subscript $t$ for the $t=3$ case. 

 Using those variables, the continuity conditions are written as 
\bea
Q^j_4(\sigma)-Q^j_4(\pi-\sigma)=0, & & P_{4,i}(\sigma)+P_{4,i}(\pi-\sigma)=0, \nn \\
Q^j_2(\sigma)+Q^j_2(\pi-\sigma)=0, & & P_{2,i}(\sigma)-P_{2,i}(\pi-\sigma)=0, \nn \\
Q^j(\sigma)-iQ^j(\pi-\sigma)=0, & & P_i(\sigma)+iP_i(\pi-\sigma)=0, \nn \\
\bar{Q}^j(\sigma)+i\bar{Q}^j(\pi-\sigma)=0, & & \bar{P}_i(\sigma)-i\bar{P}_i(\pi-\sigma)=0 
\label{conditionmodeV4}
\eea
for $0\leq\sigma\leq\frac{\pi}{2}$. In terms of the modes, 
the conditions for the sectors of $t=2$ and 4 are identical with the Neumann case 
\bea
 & & (1-C)|Q_{4,k})|\hat{V}_4\ket^X=(1+C)|P_{4,k})|\hat{V}_4\ket^X=0, \nn \\
 & & (1+C)|Q_{2,k})|\hat{V}_4\ket^X=(1-C)|P_{2,k})|\hat{V}_4\ket^X=0,  
\eea
which can be seen from the point that 
the conditions (\ref{conditionmodeV4}) for the sectors of $t=2$ and 4 lead the same 
relations between the modes as 
those without the terms containing $\theta^{jk}$. 
Here we adopted the vector notation for the modes 
\beq
|Q_{t,k})=\left[\begin{array}{c}Q_{t,0,k} \\ Q_{t,1,k} \\ \vdots \end{array}\right], 
\hspace{1cm} 
|P_{t,k})=\left[\begin{array}{c}P_{t,0,k} \\ P_{t,1,k} \\ \vdots \end{array}\right], 
\eeq
and $C$ is a matrix such that $(C)_{nm}=(-1)^n\delta_{nm}$ ($n,m\ge 0$). 
Thus there is needed no correction containing $\theta^{ij}$ for the sectors of $t=2$ and 4, 
so it is natural to assume the form of the phase factor in eq. (\ref{V4form}) as 
\beq
\frac12\theta^{ij}\sum_{r,s=1}^4(p^{(r)}_i|Z^{rs}|p^{(s)}_j)=
\theta^{ij}(P_i|Z|\bar{P}_j)
\label{phaseV4}
\eeq
with $Z$ being anti-hermitian. 

 Next let us consider the conditions for the sectors of $t=1$ and 3. 
We rewrite the mode expansions 
of $Q^j(\sigma)$ and $\bar{Q}^j(\sigma)$ as 
\bea
Q^j(\sigma) & = & G^{jk}(\sqrt{2\alpha'}Q_{0,k}
+2\sqrt{\alpha'}\sum_{n=1}^{\infty}\cos(n\sigma)Q_{n,k}) \nn \\
 & & +\theta^{jk}\left[\int_{\pi/2}^{\sigma}d\sigma'P_i(\sigma')
+\frac{1}{\pi\sqrt{\alpha'}}\sum_{n=1,3,5,\cdots}\frac1n(-1)^{(n-1)/2}P_{n,k}\right] \nn \\
 & \equiv &  \theta^{jk}\int_{\pi/2}^{\sigma}d\sigma'P_i(\sigma')+\Delta Q^j(\sigma), 
\label{modeexpansionQ}  \\
\bar{Q}^j(\sigma) & = & G^{jk}(\sqrt{2\alpha'}\bar{Q}_{0,k}+2\sqrt{\alpha'}
\sum_{n=1}^{\infty}\cos(n\sigma)\bar{Q}_{n,k}) \nn \\
 & & +\theta^{jk}\left[\int_{\pi/2}^{\sigma}d\sigma'\bar{P}_i(\sigma')
+\frac{1}{\pi\sqrt{\alpha'}}\sum_{n=1,3,5,\cdots}\frac1n(-1)^{(n-1)/2}\bar{P}_{n,k}\right]
\nn \\
 & \equiv & \theta^{jk}\int_{\pi/2}^{\sigma}d\sigma'\bar{P}_i(\sigma')+
\Delta\bar{Q}^j(\sigma) . 
\label{modeexpansionQbar}
\eea
Using the conditions for $P_i(\sigma)$ and $\bar{P}_i(\sigma)$ in (\ref{conditionmodeV4}), 
we can reduce the conditions for $Q^j(\sigma)$ and $\bar{Q}^j(\sigma)$ to 
those for $\Delta Q^j(\sigma)$ and $\Delta\bar{Q}^j(\sigma)$: 
\bea
\Delta Q^j(\sigma) & = & \left\{\begin{array}{ll}i\Delta Q^j(\pi-\sigma) & 
\hspace{1cm}(0\le \sigma\le\frac{\pi}{2}) \\ -i\Delta Q^j(\pi-\sigma) & 
\hspace{1cm}(\frac{\pi}{2}\le \sigma\le\pi),  \end{array}\right. \nn \\
\Delta\bar{Q}^j(\sigma) & = & \left\{\begin{array}{ll}-i\Delta\bar{Q}^j(\pi-\sigma) & 
\hspace{1cm}(0\le \sigma\le\frac{\pi}{2}) \\ i\Delta\bar{Q}^j(\pi-\sigma) & 
\hspace{1cm}(\frac{\pi}{2}\leq \sigma\leq\pi).  \end{array}\right. 
\eea
These formulas are translated to the relations between the modes 
via the Fourier transformation. The result is expressed in the vector notation as 
\beq
(1-X)|{\cal Q}_i)|\hat{V}_4\ket^X=(1+X)|\overline{\cal Q}_i)|\hat{V}_4\ket^X=0, 
\label{conditionQV4}
\eeq
where the matrix $X$ is defined by eqs. (\ref{defX1}) and (\ref{defX2}), also the 
vectors $|{\cal Q}_i)$ and $|\overline{\cal Q}_i)$ stand for 
\bea
|{\cal Q}_i) & = & \left[\begin{array}{c}Q_{0,i}
+\frac{i}{4\alpha'}G_{ik}\theta^{kj}\sum_{n=0}^{\infty}X_{0n}P_{n,j} \\
Q_{1,i} \\ Q_{2,i} \\ \vdots \end{array} \right],  \\
|\overline{\cal Q}_i) & = & \left[\begin{array}{c}\bar{Q}_{0,i}
+\frac{i}{4\alpha'}G_{ik}\theta^{kj}\sum_{n=0}^{\infty}X_{0n}\bar{P}_{n,j} \\
\bar{Q}_{1,i} \\ \bar{Q}_{2,i} \\ \vdots \end{array} \right]. 
\eea
In eqs. (\ref{conditionQV4}), 
passing the vectors through the phase factor of the $|\hat{V}_4\ket$ and using the 
continuity conditions in the Neumann case
\bea
 & & (1+X)|P_i)|V_4\ket^X=(1-X)|\bar{P}_i)|V_4\ket^X=0, \label{condPB=0V4} \\
 & & (1-X)|Q_i)|V_4\ket^X=(1+X)|\bar{Q}_i)|V_4\ket^X=0,
\eea 
we obtain the equations, which the coefficients $Z_{nm}$'s should satisfy,   
\bea
 & & \left[(1-X)_{m0}\sum_{n=0}^{\infty}(\bar{Z}_{0n}+i\frac{\pi}{2}\bar{X}_{0n})P_{n,j}
+\sum_{n=1}^{\infty}(1-X)_{mn}\sum_{n'=0}^{\infty}\bar{Z}_{nn'}P_{n',j}\right]|V_4\ket^X=0, \\
 & & \left[(1+X)_{m0}\sum_{n=0}^{\infty}(Z_{0n}-i\frac{\pi}{2}X_{0n})\bar{P}_{n,j}
+\sum_{n=1}^{\infty}(1+X)_{mn}\sum_{n'=0}^{\infty}Z_{nn'}\bar{P}_{n',j}\right]|V_4\ket^X=0 
\eea
for $m\geq 0$. 
Now all our remaing task is to solve these equations. 
It is easy to see that a solution of them is given by 
\bea
Z_{mn} & = & -i\frac{\pi}{2}(1-X)_{mn}+i\beta\frac{\pi}{2}C_{mn}, 
\hspace{1cm}(m,n\geq 0,  \mbox{ except the $m=n=0$ case}), \nn \\
Z_{00} & = & i\beta\frac{\pi}{2}, 
\eea
if we pay attention to eqs. (\ref{condPB=0V4}). 
Here $\beta$ is an unknown real constant, which is not fixed by the continuity conditions 
alone. This ambiguity of the solution comes from the property 
of the matrix $X$: $XC=-CX$. However it will become clear 
that the term containing the constant $\beta$ does not contribute 
to the vertex $|\hat{V}_4\ket^X$. 

Therefore, we have the expression of the phase (\ref{phaseV4}) 
\beq
\theta^{ij}(P_i|Z|\bar{P}_j)=
\theta^{ij}\left[i\frac{\pi}{2}P_{0,i}\bar{P}_{0,j}
+i\beta\frac{\pi}{2}\sum_{n=0}^{\infty}(-1)^nP_{n,i}\bar{P}_{n,j}
-i\frac{\pi}{2}\sum_{m,n=0}^{\infty}P_{m,i}(1-X)_{mn}\bar{P}_{n,j}\right]. 
\eeq
Then recalling the eqs. (\ref{condPB=0V4}) again, the last term in the r. h. s. can be 
discarded.  Also we can rewrite the term containing $\beta$ 
\bea
-\frac{\theta^{ij}}{4\alpha'}\beta(P_i|C|\bar{P}_j) & = & 
+\frac{\theta^{ij}}{4\alpha'}\beta(P_i|X^TCX|\bar{P}_j) \nn \\
 & = & +\frac{\theta^{ij}}{4\alpha'}\beta(P_i|C|\bar{P}_j), 
\eea
on $|V_4\ket^X$. When going to the second line, 
the properties of the matrix $X$ (\ref{propertiesX}) were used. 
The above formula means that the term containing $\beta$ can be set to 
zero on $|V_4\ket^X$.
After all, the form of the 4-string vertex becomes 
\beq
|\hat{V}_4\ket^X_{1234}=\exp\left[-\frac{\theta^{ij}}{4\alpha'}P_{0,i}\bar{P}_{0,j}
\right]|V_4\ket^X_{1234}. 
\eeq
Note that the phase factor has the cyclic symmetric form  
\beq
-\frac{\theta^{ij}}{4\alpha'}P_{0,i}\bar{P}_{0,j}
= i\frac{\theta^{ij}}{8\alpha'}(p^{(1)}_{0.i}p^{(2)}_{0,j}+p^{(2)}_{0.i}p^{(3)}_{0,j}+
p^{(3)}_{0.i}p^{(4)}_{0,j}+p^{(4)}_{0.i}p^{(1)}_{0,j}),  
\eeq
which is a property the vertices should have\footnote{Here 
the momentum $p^{(r)}_{0,i}$ is 
given by $p^{(r)}_{0,i}=\sqrt{2\alpha'}p^{(r)}_i$.}.

\subsection{$|\hat{V}_3\ket^X_{123}$}

   We can obtain the 3-string overlap in the similar manner as in the 4-string case. 
First, we introduce the $Z_3$-Fourier transformed variables 
\bea
Q^j_1(\sigma)  & = & 
\frac{1}{\sqrt{3}}[eX^{(1)j}(\sigma)+\bar{e}X^{(2)j}(\sigma)+X^{(3)j}(\sigma)]  
\equiv Q^j(\sigma), \nn \\
Q^j_2(\sigma)  & = & 
\frac{1}{\sqrt{3}}[\bar{e}X^{(1)j}(\sigma)+eX^{(2)j}(\sigma)+X^{(3)j}(\sigma)] 
\equiv \bar{Q}^j(\sigma), \nn \\
Q^j_3(\sigma) & = & \frac{1}{\sqrt{3}}[X^{(1)j}(\sigma)+X^{(2)j}(\sigma)+X^{(3)j}(\sigma)], 
\eea
where $e\equiv e^{i2\pi/3}$, $\bar{e}\equiv e^{-i2\pi/3}$. The momenta 
$P_{1,i}(\sigma)(\equiv P_i(\sigma))$, $P_{2,i}(\sigma)(\equiv \bar{P}_i(\sigma))$ and 
$P_{3,i}(\sigma)$ are defined in the same way. 
The mode expansions take the same form as those in eqs. (\ref{modeexpansionZ4}). 
In these variables, the continuity conditions require 
\bea
Q^j(\sigma)-eQ^j(\pi-\sigma)=0, & & 
P_i(\sigma)+eP_i(\pi-\sigma)=0, \nn \\
\bar{Q}^j(\sigma)-\bar{e}\bar{Q}^j(\pi-\sigma)=0, & & 
\bar{P}_i(\sigma)+\bar{e}\bar{P}_i(\pi-\sigma)=0, \nn \\
Q^j_3(\sigma)-Q^j_3(\pi-\sigma)=0, & & 
P_{3,i}(\sigma)+P_{3,i}(\pi-\sigma)=0
\eea
for $0\leq\sigma\leq\frac{\pi}{2}$. The conditions imposed to the modes 
with respect to the $t=3$ component are identical with those in the Neumann case 
\beq
(1+C)|P_{3,i})|\hat{V}_3\ket^X=(1-C)|Q_{3,i})|\hat{V}_3\ket^X=0. 
\eeq
Thus the $t=3$ component does not couple with $\theta^{ij}$, 
so we can find 
the solution by determining the single anti-hermitian matrix $Z$ 
in the phase factor whose form is assumed as 
\beq
\frac12\theta^{ij}\sum_{r,s=1}^3(p^{(r)}_i|Z^{rs}|p^{(s)}_j)=
\theta^{ij}(P_i|Z|\bar{P}_j). 
\label{phaseV3}
\eeq

For the sectors of $t=1$ and 2, the same argument goes on as in the 4-string case. 
$Q^j(\sigma)$ and $\bar{Q}^j(\sigma)$ have the mode expansions same as in 
eqs. (\ref{modeexpansionQ}) and (\ref{modeexpansionQbar}). 
The conditions we have to consider are 
\bea
\Delta Q^j(\sigma) & = & \left\{\begin{array}{ll}e\Delta Q^j(\pi-\sigma) & 
\hspace{1cm}(0\le \sigma\le\frac{\pi}{2}) \\ \bar{e}\Delta Q^j(\pi-\sigma) & 
\hspace{1cm}(\frac{\pi}{2}\le \sigma\le\pi),  \end{array}\right. \nn \\
\Delta\bar{Q}^j(\sigma) & = & \left\{\begin{array}{ll}\bar{e}\Delta\bar{Q}^j(\pi-\sigma) & 
\hspace{1cm}(0\le \sigma\le\frac{\pi}{2}) \\ e\Delta\bar{Q}^j(\pi-\sigma) & 
\hspace{1cm}(\frac{\pi}{2}\leq \sigma\leq\pi),   \end{array}\right. 
\eea
which are rewritten as the relations between the modes 
\beq
(1-Y)|{\cal Q}_i)|\hat{V}_3\ket^X=(1-Y^T)|\overline{\cal Q}_i)|\hat{V}_3\ket^X=0. 
\label{conditionQV3}
\eeq
The matrix $Y$ is defined below eqs. (\ref{equationsU}). 
Recalling the conditions in the Neumann case 
\bea
 & & (1+Y)|P_i)|V_3\ket^X=(1+Y^T)|\bar{P}_i)|V_3\ket^X=0,  
\label{conditionPV3} \\
 & & (1-Y)|Q_i)|V_3\ket^X=(1-Y^T)|\bar{Q}_i)|V_3\ket^X=0, 
\eea
we end up with the following equations 
\bea
 & & \left[(1-Y)_{m0}\sum_{n=0}^{\infty}(\bar{Z}_{0n}+\frac{\pi}{2}\bar{X}_{0n})P_{n,j}
+\sum_{n=1}^{\infty}(1-Y)_{mn}\sum_{n'=0}^{\infty}\bar{Z}_{nn'}P_{n',j}\right]|V_3\ket^X=0, 
\\
 & & \left[(1-Y^T)_{m0}\sum_{n=0}^{\infty}(Z_{0n}-i\frac{\pi}{2}X_{0n})\bar{P}_{n,j} 
+\sum_{n=1}^{\infty}(1-Y^T)_{mn}\sum_{n'=0}^{\infty}Z_{nn'}\bar{P}_{n',j}\right]|V_3\ket^X=0 
\eea
for $m\geq 0$. It can be easily found out that the expression 
\bea
Z_{mn} & = & -i\frac{\pi}{\sqrt{3}}(1+Y^T)_{mn} \hspace{1cm} 
(m,n\geq 0, \mbox{ except the $m=n=0$ case}), \nn \\
Z_{00} & = & 0
\eea
satisfies the above equations. It should be noted that in this case, 
because of $CYC=\bar{Y}\neq -Y$, it does not 
contain any unknown constant differently from the 4-string case. 

 Owing to the condition (\ref{conditionPV3}) we can write the phase factor only 
in terms of the zero-modes. Finally we have 
\bea
|\hat{V}_3\ket^X_{123} & = & 
\exp\left[-\frac{\theta^{ij}}{4\sqrt{3}\alpha'}P_{0,i}\bar{P}_{0,j}\right]
|V_3\ket^X_{123}
\nn \\
 & = & \exp\left[i\frac{\theta^{ij}}{12\alpha'}
(p^{(1)}_{0,i}p^{(2)}_{0,j}+p^{(2)}_{0,i}p^{(3)}_{0,j}+p^{(3)}_{0,i}p^{(1)}_{0,j})
\right]|V_3\ket^X_{123}.
\label{overlapV3}
\eea

 It is not clear whether the solutions we have obtained here are unique or not. 
However we can show that the phase factors are consistent with the relations between the 
overlaps which they should satisfy : 
\beq
_3\!\bra \hat{I}|\hat{V}_3\ket_{123}=|\hat{V}_2\ket_{12}, \hspace{1cm}
_4\!\bra \hat{I}|\hat{V}_4\ket_{1234}=|\hat{V}_3\ket_{123}, \hspace{1cm}
_{34}\!\bra \hat{V}_2||\hat{V}_3\ket_{123}|\hat{V}_3\ket_{456}=|\hat{V}_4\ket_{1256}, 
\eeq
by using the momentum conservation on the vertices 
$(p^{(1)}_i+\cdots+p^{(N)}_i)|\hat{V}_N\ket^X_{1\cdots N}=0$. 
Furthermore we can see that the phase factors successfully 
reproduce the Moyal product structures of the 
correlators among vertex operators obtained in the perturbative approach 
to open string theory 
in the constant $B$-field background \cite{SW}. 
These facts convince us that the solutions obtained here are  
physically meaningful.

%%%%%%%%%%% Section 5: Transformation of String Fields  %%%%%%%%%%%%%%%%%%%%%%

\section{Transformation of String Fields}
\setcounter{equation}{0}
   In the previous section, we have explicitly constructed the overlap vertices 
in the operator formulation under the constant $B$-field background. 
Then we have obtained the vertices with a new noncommutative structure of the 
Moyal type originating from the constant $B$-field, in addition to 
the ordinary product $*$ of string fields. 
Denoting the product with the new structure by $\star$, the action of the 
string field theory is written as 
\bea
S_B & = & \frac{1}{G_s}\int \left(\frac12 \psi \star Q\psi+
\frac13 \psi\star\psi\star\psi\right)      \nn \\
 & = & \frac{1}{G_s}
\left(\frac12 \mbox{ }_{12}\!\bra \hat{V}_2||\psi\ket_1 Q|\psi\ket_2
+\frac13 \mbox{ }_{123}\!\bra \hat{V}_3||\psi\ket_1 |\psi\ket_2 |\psi\ket_3\right), 
\label{SFTactionB1}
\eea
where the BRST charge $Q$ is constructed from the world sheet action (\ref{wsactionB}). 
The theory (\ref{SFTactionB1}) gives the noncommutative $U(1)$ Yang-Mills theory 
in the low energy region 
in the same sense as Witten's open string field theory 
in the case of the Neumann boundary condition leads to the ordinary $U(1)$ Yang-Mills 
theory in the low energy limit\footnote{It can be explicitly seen by repeating a 
similar calculation as that carried out in ref. \cite{Dearnaley2}. }.

  In ref. \cite{SW}, Seiberg and Witten argued that open string theory 
in the constant $B$-field background leads to either commutative or noncommutative 
Yang-Mills theories, corresponding to the different regularization scheme ---
the Pauli-Villars regularization or the point-splitting regularization --- 
in the world sheet formulation. They discussed a map between the gauge fields in 
the commutative and noncommutative Yang-Mills theories. 
In string field theory perspective, there also should be a certain transformation 
(hopefully simpler than the Yang-Mills case) from the string field $\psi$ in (\ref{SFTactionB1}) 
to a string field in a new string field theory 
which leads to the commutative Yang-Mills theory in the low energy limit. 

  Here we obtain the new string field theory by finding a unitary transformation 
which absorbs the noncommutative structure of the Moyal type 
in the product $\star$ into a redefinition of the string fields. 
There are used the two vertices $|\hat{V}_2\ket$ and $|\hat{V}_3\ket$ 
in the action (\ref{SFTactionB1}). 
Recall that the 2-string vertex is in the same form as 
in the Neumann case and has no Moyal type noncommutative structure. 
First, we consider the phase factor of the 3-string vertex which multiplies in front of 
$|V_3\ket$ (See eq. (\ref{overlapV3}).). Making use of the continuity conditions 
\beq
P_{0,i}=-2\sum_{n=1}^{\infty}Y_{0n}P_{n,i}, \hspace{1cm}
\bar{P}_{0,i}=-2\sum_{n=1}^{\infty}\bar{Y}_{0n}\bar{P}_{n,i},
\label{condP}
\eeq
it can be rewritten as 
\bea
-\frac{\theta^{ij}}{4\sqrt{3}\alpha'}P_{0,i}\bar{P}_{0,j} 
 & = & \frac{\theta^{ij}}{4\sqrt{3}\alpha'}\sum_{n=1}^{\infty}
(P_{0,i}\bar{Y}_{0n}\bar{P}_{n,j}+P_{n,i}Y_{0n}\bar{P}_{0,j}) \nn \\
 & = &  -\frac{\theta^{ij}}{24\alpha'}\sum_{n=1}^{\infty}X_{0n}
[(-p^{(2)}_{0,i}-p^{(3)}_{0,i}+2p^{(1)}_{0,i})p^{(1)}_{n,j} \nn \\
 & & \hspace{1cm} +(-p^{(3)}_{0,i}-p^{(1)}_{0,i}+2p^{(2)}_{0,i})p^{(2)}_{n,j}
+(-p^{(1)}_{0,i}-p^{(2)}_{0,i}+2p^{(3)}_{0,i})p^{(3)}_{n,j}] \nn \\
 & = & -\frac{\theta^{ij}}{8\alpha'}\sum_{r=1}^3\sum_{n=1}^{\infty}X_{0n}
p^{(r)}_{0,i}p^{(r)}_{n,j},
\eea
where we used the property of the matrix $Y$: 
$Y_{0n}=-\bar{Y}_{0n}=\frac{\sqrt{3}}{2}X_{0n}$ for $n\geq 1$ 
and the momentum conservation 
on $|V_3\ket$: $p^{(1)}_{0,i}+p^{(2)}_{0,i}+p^{(3)}_{0,i}=0$. 
We manage to represent the phase factor of the Moyal type as a form factorized into 
the product of the unitary operators 
\beq
{{\cal U}}_r=\exp\left(\frac{\theta^{ij}}{8\alpha'}\sum_{n=1,3,5,\cdots}
X_{0n}p^{(r)}_{0,i}p^{(r)}_{n,j}\right). 
\label{U}
\eeq
Note that the unitary operator acts to a single string field. 
So 
the Moyal type noncommutativity can be absorbed by the unitary rotation of the 
string field 
\bea
_{123}\!\bra \hat{V}_3||\psi\ket_1 |\psi\ket_2 |\psi\ket_3 & = & 
_{123}\!\bra V_3|{{\cal U}}_1{{\cal U}}_2{{\cal U}}_3
|\psi\ket_1 |\psi\ket_2 |\psi\ket_3 \nn \\
 & = & _{123}\!\bra V_3||\tilde{\psi}\ket_1|\tilde{\psi}\ket_2|\tilde{\psi}\ket_3, 
\label{manipulationV3}
\eea
with $|\tilde{\psi}\ket_r={{\cal U}}_r|\psi\ket_r$. 
It should be remarked that this manipulation has been suceeded 
owing to the factorized expression of 
the phase factor, which originates from the continuity 
conditions relating the zero-modes to the nonzero-modes (\ref{condP}). 
It is a characteristic 
feature of string field theory that can not be found in any local field theories. 

Next let us see the kinetic term. 
In doing so, it is judicious to write the kinetic term as follows: 
\beq
_{12}\!\bra\hat{V}_2||\psi\ket_1(Q|\psi\ket_2)=
_{123}\!\bra\hat{V}_3||\psi\ket_1(Q_L|I\ket_2|\psi\ket_3+|\psi\ket_2Q_L|I\ket_3), 
\label{rewrittenkin1}
\eeq
where $Q_L$ is defined by integrating the BRST current $j_{BRST}(\sigma)$ 
with respect to $\sigma$ over 
the left half region 
\beq
Q_L=\int_0^{\pi/2}d\sigma j_{BRST}(\sigma). 
\eeq
Eq. (\ref{rewrittenkin1}) is also represented by the product $\star$ as 
\beq
\psi\star (Q\psi)=\psi\star [(Q_LI)\star\psi+\psi\star(Q_LI)]. 
\label{rewrittenkin2}
\eeq
$I$ stands for the identity element with respect to the $\star$-product, 
carrying the ghost number $-\frac32$, 
which corresponds to $|I\ket$ in the operator formulation. 
As is discussed in ref. \cite{HLRS}, in order to show the relation (\ref{rewrittenkin2}) 
we need the formulas 
\bea
 & & Q_RI=-Q_LI,  \nn  \\
 & & (Q_R\psi)\star\xi=-(-1)^{n_{\psi}}\psi\star(Q_L\xi) 
\label{QLQR}
\eea
for arbitrary string fields $\psi$ and $\xi$, where  
$Q_R$ is the integrated BRST current over the right half region of $\sigma$. 
$n_{\psi}$ stands for the ghost number of the string field $\psi$ minus $\frac12$, 
and takes an integer value. 
The first formula means that the identity element is a physical quantity, also 
the second does the conservation of the BRST charge\footnote{For a proof of 
these formulas, we can use the same argument as in 
ref. \cite{HLRS} and references therein.}. 
By using these formulas, 
the first term in the bracket in r. h. s. of eq. (\ref{rewrittenkin2}) becomes 
$$
(Q_LI)\star\psi=-(Q_RI)\star\psi=I\star(Q_L\psi)=Q_L\psi. 
$$
Also, it turns out that the second term is equal to $Q_R\psi$. 
Combining these, we can see 
that the eq. (\ref{rewrittenkin2}) holds. 

Further, we should remark that because the BRST current does not contain 
the center of mass coordinate $x^j$, 
it commute with the momentum $p_i$. From the continuity condition $p_i|I\ket=0$, 
it can be seen that $p_iQ_L|I\ket=0$. 
Expanding the exponential in the expression of the unitary operator (\ref{U}) and 
passing the momentum $p_{0,i}$ to the right, we obtain 
\beq
{{\cal U}}Q_L|I\ket=Q_L|I\ket. 
\label{UQL}
\eeq
Now we can write down the result of the kinetic term.  
As a result of the same manipulation as in eq. (\ref{manipulationV3}) 
and the use of eq. (\ref{UQL}), we have\footnote{Strictly speaking, 
in general this formula holds in the case that both of the string fields 
$|\psi\ket$ and $|\tilde{\psi}\ket$ belong to the Fock space which consists of states 
excited by {\it finite} number of creation operators. This point is subtle for giving a 
proof. However, for the infinitesimal $\theta$ case, by keeping arbitrary finite order terms 
in the expanded form of the exponential of ${{\cal U}}$, we can make the situation of 
both $|\psi\ket$ and $|\tilde{\psi}\ket$ being inside the Fock space, and thus clearly 
eq. (\ref{subtleformula}) holds. From this fact, it is plausible to expect that 
eq. (\ref{subtleformula}) is correct in the finite $\theta$ case.}  
\bea
_{12}\!\bra\hat{V}_2||\psi\ket_1(Q|\psi\ket_2) & = & 
_{123}\!\bra\hat{V}_3||\psi\ket_1(Q_L|I\ket_2|\psi\ket_3+|\psi\ket_2Q_L|I\ket_3) \nn \\
 & = & 
_{123}\!\bra V_3||\tilde{\psi}\ket_1(Q_L|I\ket_2|\tilde{\psi}\ket_3+
|\tilde{\psi}\ket_2Q_L|I\ket_3)
\nn \\
 & = & 
_{12}\!\bra V_2||\tilde{\psi}\ket_1(Q|\tilde{\psi}\ket_2). 
\label{subtleformula}
\eea

Here we have a comment. 
If we considered the kinetic term itself without using eq. (\ref{rewrittenkin1}), 
what would be going on? 
Let us see this. From the continuity conditions for 
$|\hat{V}_2\ket^X_{12}=|V_2\ket^X_{12}$: 
\beq
p^{(1)}_{0,i}+p^{(2)}_{0,i}=0, 
\hspace{1.5cm}p^{(1)}_{n,i}+(-1)^np^{(2)}_{n,i}=0 \hspace{1cm}(n=1,2,\cdots), 
\eeq
it could be shown that the 2-string overlap is invariant under the unitary rotation 
\beq
{{\cal U}}_1{{\cal U}}_2|V_2\ket_{12}=|V_2\ket_{12}.
\eeq
So we would find the expression for the kinetic term after the rotation 
\beq
_{12}\!\bra V_2||\psi\ket_1Q|\psi\ket_2 
=\mbox{ }_{12}\!\bra V_2||\tilde{\psi}\ket_1\tilde{Q}|\tilde{\psi}\ket_2, 
\eeq
where $\tilde{Q}$ is the BRST charge similarity transformed by ${{\cal U}}$
\beq
\tilde{Q}={{\cal U}}Q{{\cal U}}^{\dagger}. 
\label{Qtilde}
\eeq
However, after some computations of the r. h. s. of eq. (\ref{Qtilde}), 
we could see that $\tilde{Q}$ has divergent term proportional to 
$$
\sum_{n=1,3,5,\cdots}1  
$$
and thus it is not well-defined. 
It seems that this procedure is not correct and needs 
some suitable regularization, which preserves the conformal 
symmetry\footnote{That divergence comes from the mid-point 
sigularity of the string coordinates transformed by ${{\cal U}}$. 
In fact, after some calculations, we have 
\beq
{{\cal U}}X^j(\sigma){{\cal U}}^{\dagger}
=X^j(\sigma)-i\frac{\theta^{jk}}{4\sqrt{2\alpha'}}
\sum_{n=1,3,5,\cdots}X_{n0}p_{n,k}-
\frac{\theta^{jk}}{4}p_k\mbox{ sgn}\left(\sigma-\frac{\pi}{2}\right). 
\label{transformedcoordinate}
\eeq
The last term leads to the mid-point sigularity in the energy-momentum tensor and the 
BRST charge $Q$. 
It seems that the use of eq. (\ref{rewrittenkin1}) corresponds to taking the 
point splitting regularization with respect to the mid-point. 
Because of the discontinuity of the last term in eq. (\ref{transformedcoordinate}), 
it is considered that the transformed string coordinates have no longer a good 
picture as a string. 
It could be understood from the point that the transformation ${{\cal U}}$ drives 
states around a perturbative vacuum to those around highly nonperturbative one 
like coherent states.}. 
It is considered 
that the use of eq. (\ref{rewrittenkin1}) gives 
that kind of regularization, 
which will be justified at the end of the next section. 

  Therefore the string field theory action (\ref{SFTactionB1}) 
with the Moyal type noncommutativity added to the ordinary noncommutativity 
is equivalently 
rewritten as the one with the ordinary noncommutativity alone: 
\bea
S_B & = & \frac{1}{G_s}\int \left(\frac12 \tilde{\psi} * Q\tilde{\psi}
+\frac13 \tilde{\psi}*\tilde{\psi}*\tilde{\psi}\right)      \nn \\
 & = & \frac{1}{G_s}
\left(\frac12 \mbox{ }_{12}\!\bra V_2||\tilde{\psi}\ket_1 Q|\tilde{\psi}\ket_2
+\frac13 \mbox{ }_{123}\!\bra V_3||\tilde{\psi}\ket_1 |\tilde{\psi}\ket_2 
|\tilde{\psi}\ket_3\right). 
\label{SFTactionB2}
\eea
It is noted that the BRST charge $Q$, which is constructed from the world sheet action 
(\ref{wsactionB}), has the same form as the one obtained from the action (\ref{wsactionG}) 
with the relation (\ref{G}). So all the $B$-dependence has been stuffed into 
the string fields except that existing in the metric $G_{ij}$. 
Furthermore, recalling that the relation between the metrics $G^{ij}$ and $g_{ij}$ is 
the same form as the T-duality inversion transformation, 
which was pointed out at the end of section 3, we can make the metric $g_{ij}$ 
appear in the overlap vertices, instead of the metric $G_{ij}$. 
To do so, we consider the following transformation for the modes: 
\beq
\hat{\alpha}^i_n=(E^{T-1})^{ik}\alpha_{n,k}, 
\hspace{1cm} \hat{p}^i=(E^{T-1})^{ik}p_k, \hspace{1cm} \hat{x}_i=E_{ik}x^k.
\label{Ttransf}
\eeq
By this transformation, the commutators become 
\beq
[\hat{\alpha}^i_n, \hat{\alpha}^j_m]=ng^{ij}\delta_{n+m,0}, 
\hspace{1cm} [\hat{p}^i, \hat{x}_j]=-i\delta^i_j, 
\eeq
and the bilinear form of the modes 
\bea
G^{ij}\alpha_{n,i}\alpha_{m,j}  & =  & g_{ij}\hat{\alpha}^i_n\hat{\alpha}^j_m, \nn \\
G^{ij}p_i\alpha_{m,j}& =  & g_{ij}\hat{p}^i\hat{\alpha}^j_m, \nn \\
G^{ij}p_ip_j & = & g_{ij}\hat{p}^i\hat{p}^j. 
\label{bilinear}
\eea
Since in viewing the expressions of the overlaps (\ref{V2B=0}) and (\ref{V3B=0}) 
the metric $G_{ij}$ appears only through the bilinear forms 
(\ref{bilinear}), now the dependence 
of the background $B_{ij}$ can be completely 
removed from the overlaps $|V_2\ket$ and $|V_3\ket$ in 
the action (\ref{SFTactionB2}) after the transformation (\ref{Ttransf}). 

Remark that we have performed the unitary rotation and the T-duality inversion 
but have not any shift of the string field. 
In string field theory perspective, 
the background is changed by shifting the string field by a solution of the classical 
string field equation. 
Because we have done no such a shift, 
it is possible to consider that 
the two actions (\ref{SFTactionB1}) and (\ref{SFTactionB2}) 
describe open strings in the same background.  
The one (\ref{SFTactionB1}) certainly leads to the noncommutative Yang-Mills theory in the 
low energy limit. 
Also, it is clear 
that the other (\ref{SFTactionB2}) gives the commutative Yang-Mills theory as the low energy 
effective theory. Thus it is considered that the unitary transformation 
$|\tilde{\psi}\ket={{\cal U}}|\psi\ket$ combined with 
the T-duality inversion (\ref{Ttransf}) corresponds to a 
microscopic analog of the map between the gauge fields 
in the commutative and noncommutative 
Yang-Mills theories in ref. \cite{SW}. 

We end this section by noting a following comment. 
We have had the string field theories which describe the commutaive and noncommutative 
Yang-Mills theories at the low energy region. They have the identical string coupling 
$G_s$. On the other hand, according to an analysis of the Dirac-Born-Infeld (DBI) actions 
corresponding to the commutative and noncommutative Yang-Mills theories in ref. \cite{SW}, 
it turns out that the coupling constants of the commutative and noncommutative DBI actions, 
denoted by $g_s^{(DBI)}$ and $G_s^{(DBI)}$ respectively, differ by a factor 
depending on the metrics: 
\beq
G^{(DBI)}_s(\det G)^{-1/4}=g_s^{(DBI)}(\det g)^{-1/4}. 
\label{SWresult}
\eeq
This relation seems to be nontrivial from our point of view. 
In the commutative theory we have had, since all the $B$-dependence is contained 
in infinite number of wave functions in the string fields, it seems to 
need more consideration to 
derive eq. ({\ref{SWresult}). We will discuss this issue elsewhere. 

%At a glance, our result seems not to be consistent with (\ref{SWresult}). 
%However, here we should remark that the string coupling $G_s$ is not identical with the 
%coupling appearing in the DBI action $G_s^{(DBI)}$. They are related by 
%\beq
%G_s^{(DBI)}=G_s(\det G)^{1/4}, 
%\eeq
%which can be seen in the process of a calculation of the disk amplitude\footnote{For 
%example, see ref. \cite{Lee}.}.
%Thus, our result matches with the analysis of the DBI actions. 

%%%%%%%%%%%%%%%%%%%%%% Section 6: Background Independence %%%%%%%%%%%%%%%%%%%%%%%%%%%%%%%%
\section{Background Independence}
\setcounter{equation}{0}
  In this section, 
we show that both theories in the background described by the world sheet 
action (\ref{wsactionB}) and in the background of the metric 
$g_{ij}$ alone are derived from the single purely cubic action 
\beq
S_3=\frac{1}{3G_s}\int\psi\star\psi\star\psi=
\frac{1}{3G_s}\mbox{ }_{123}\!\bra \hat{V}_3||\psi\ket_1 |\psi\ket_2 |\psi\ket_3, 
\label{cubicaction}
\eeq
by using the unitary transformation (\ref{U}) together with an argument in 
ref. \cite{HLRS}. The result gives a new type of the background independence connecting 
between theories with different boundary conditions caused by the background $B_{ij}$. 

First we note that the cubic action has a universal structure. 
The Moyal type noncommutativity is absorbed into the redefinition of the string field 
$|\tilde{\psi}\ket={{\cal U}}|\psi\ket$ as discussed in the previous section. 
Also we can get rid of the dependence of the 
metric $G_{ij}$. Since $G_{ij}$ can be diagonalized by an appropriate orthogonal matrix, 
we rotate the modes $\alpha_{n,i}$, $p_i$ and $x^i$ by the orthogonal matrix and 
absorb the eigenvalues into suitable rescalings of the modes. 
So the metric appearing in the commutators and in the bilinear forms 
can be made Minkowskian: 
\bea
[\alpha_{n,i}, \alpha_{m,j}]=nG_{ij}\delta_{n+m,0}  & \Rightarrow & 
[\alpha_{n,i}, \alpha_{m,j}]=n\eta_{ij}\delta_{n+m,0},  \nn \\
G^{ij}\alpha_{n,i}\alpha_{m,j}   & \Rightarrow & \eta^{ij}\alpha_{n,i}\alpha_{m,j}, \nn \\
G^{ij}p_i\alpha_{m,j}   & \Rightarrow & \eta^{ij}p_i\alpha_{m,j}, \nn \\
G^{ij}p_ip_j & \Rightarrow & \eta^{ij}p_ip_j. 
\eea
As a result, the cubic action (\ref{cubicaction}) is to 
have no specific dependence of the background $g_{ij}$ and $B_{ij}$.  
It shows that the universality of the interaction vertices in string field theory 
discussed in refs. \cite{Kugo-Zwiebach,Yoneya,HLRS} holds also 
between the open string field theories with different boundary conditions. 

The classical equation of motion obtained from the action (\ref{cubicaction}) is 
\beq
\psi\star\psi=0. 
\label{eqofmotion}
\eeq
Let us denote the BRST charge derived from the world sheet action (\ref{wsactionB}) by $Q$. 
We see that $\psi=Q_LI$ gives a solution of (\ref{eqofmotion}) by using the properties 
\beq
\{Q,Q_L\}=0 
\label{QQL}
\eeq
in addition to eqs. (\ref{QLQR}). 
When showing eq. (\ref{QQL}) by a direct calculation, 
it is crucial that the BRST current $j_{BRST}$ 
is a primary field with the conformal weight 1. 
We decompose the string field $\psi$ into the solution $Q_LI$ and the fluctuation around it 
$\Psi$. Then, we can rewrite the cubic action into Witten's form (\ref{SFTactionB1}) 
by using eq. (\ref{rewrittenkin1}). 

On the other hand, via the unitary transformation (\ref{U}), 
the cubic action (\ref{cubicaction}) becomes 
\beq
S_3=\frac{1}{3G_s}\int\tilde{\psi}*\tilde{\psi}*\tilde{\psi}=
\frac{1}{3G_s}\mbox{ }_{123}\!\bra V_3|
|\tilde{\psi}\ket_1 |\tilde{\psi}\ket_2 |\tilde{\psi}\ket_3.   
\label{cubicaction2}
\eeq
We write as $Q^0$ 
the BRST charge corresponding to the world sheet action in the background 
of the metric $g_{ij}$ alone. 
After considering the transformation (\ref{Ttransf}), it is found that 
$\tilde{\psi}=Q^0_LI$ satisfies the equation of motion $\tilde{\psi}*\tilde{\psi}=0$. 
We expand the string field around this solution as 
$\tilde{\psi}=Q^0_LI+\tilde{\Psi}$, and then 
the string field action in the background of the metric alone is obtained: 
\beq
S^0=\frac{1}{G_s}\int\left(\frac12\tilde{\Psi}*(Q^0\tilde{\Psi})
+\frac13\tilde{\Psi}*\tilde{\Psi}*\tilde{\Psi}\right). 
\label{SFTaction0}
\eeq

By combining these two facts we conclude 
that the theory in the background $g_{ij}$ and $B_{ij}$ (\ref{SFTactionB1}) 
is connected to the theory in the background $g_{ij}$ alone (\ref{SFTaction0}). 
It is the advertised result. 

At the end of this section we give a justification of the regularization used in the 
previous section by showing the equivalence of the actions (\ref{SFTactionB1}) 
and (\ref{SFTactionB2}). If we do not consider the transformation (\ref{Ttransf}) 
in the action (\ref{cubicaction2}), we naturally find $\tilde{\psi}=Q_LI$ as a solution. 
In fact, because both BRST currents corresponding to $Q^0$ and $Q$ are primary fields 
with the conformal weight 1, they are qualified to be a background of string theory. 
After the same procedure as the above, we have the action 
\beq
S=\frac{1}{G_s}\int\left(\frac12\tilde{\Psi}*(Q\tilde{\Psi})
+\frac13\tilde{\Psi}*\tilde{\Psi}*\tilde{\Psi}\right),  
\label{SFTactionB3}
\eeq
which represents the theory in the background $g_{ij}$ and $B_{ij}$. 
Thus it turns out that the actions (\ref{SFTactionB1}) and (\ref{SFTactionB3}) 
describe the theory in the same background, which is nothing but the equivalence of 
eq. (\ref{SFTactionB1}) and eq. (\ref{SFTactionB2}).

%%%%%%%%%%%%%%%%%%%%%% Section 7: Discussions %%%%%%%%%%%%%%%%%%%%%%%%%%%%%%%%
\section{Discussions}
\setcounter{equation}{0}
  We have considered Witten's open string field theory in the presence of 
a constant background of the second-rank antisymmetric tensor field $B_{ij}$. 
We have extended the operator formulation based on a Fock space representation 
of string fields and overlap vertices in this situation, and have constructed 
explicitly the overlap vertices. 

  As a result, we have found a new kind of noncommutative structure 
in addition to the ordinary noncommutativity which exists 
irrespective of the $B$-field background. 
It appears only in a part of the string field product, 
which is the part concerning the zero-modes, 
and the form is a product of the Moyal type. 
It is consistent with the result 
of the correlation functions among vertex operators in the world sheet formulation. 
The string field theory obtained here gives the noncommutative Yang-Mills theory 
in the low energy limit. 

 Furthermore we have found that the Moyal type noncommutativity can be absorbed 
into a redefinition of the string fields by a unitary transformation. 
After the transformation, the string field 
theory is rewritten into a string field theory with the ordinary noncommutative product. 
Although there still exists the $B$-field dependence in the metric appearing in the 
overlaps, we can perform the T-duality inversion transformation 
so as to completely absorb it. 
This theory describes the commutative Yang-Mills theory in the low energy region. 
The unitary transformation plus the T-duality inversion transformation is considered to be 
a microscopic analog of the map between the gauge fields 
in the commutative and noncommutative 
Yang-Mills theories discussed in ref. \cite{SW}. 

 Also we have considered the background independence in this situation along the same 
line as discussed in ref. \cite{HLRS}. It has been shown that the theory in the background 
of the $B$-field as well as the metric is connected to that in the background of 
the metric alone by an appropriate redefinition of the string fields. 
We have seen that the universality of the interaction vertices in string field theory 
argued before in refs. \cite{Kugo-Zwiebach,Yoneya,HLRS} holds also between 
the open string theories with different boundary conditons 
caused by the $B$-field background. 

 Until now, we have not put the Chan-Paton factors to the end points of open strings, 
which corresponds to considering the case of the $U(1)$-gauge group in Yang-Mills theory. 
However, as is pointed out in refs. \cite{Dearnaley1,Dearnaley2}, it is a straghtforward 
generalization to introduce the Chan-Paton factors in 
Witten's open string field theory: 
\bea
S^{CP} & = & \frac{1}{G_s}\int \left(\frac12 \psi_a *Q\psi_b\:\tr(\lambda_a\lambda_b)
+\frac13 \psi_a*\psi_b*\psi_c\:\tr(\lambda_a\lambda_b\lambda_c)\right) \nn \\
 & = & \frac{1}{G_s}\left(\frac12 \mbox{ }_{12}\!\bra V_2||\psi_a\ket_1 Q|\psi_b\ket_2
\:\tr(\lambda_a\lambda_b)
+\frac13 \mbox{ }_{123}\!\bra V_3||\psi_a\ket_1 |\psi_b\ket_2 |\psi_c\ket_3
\:\tr(\lambda_a\lambda_b\lambda_c)\right), 
\label{SFTactionCP}
\eea
where $a$, $b$ and $c$ stand for the Chan-Paton factors, and $\lambda_a$ is a generator 
of the gauge group. Since the structure of the gauge group and 
that of the overlaps appear in the factorized form, 
we can consider them independently. 
Thus we can immediately write down the string field theory action 
with the Chan-Paton factors 
in the presence of the $B$-field background as 
\bea
S^{CP}_B & = & 
\frac{1}{G_s}\int \left(\frac12 \psi_a \star Q\psi_b\:\tr(\lambda_a\lambda_b)
+\frac13 \psi_a\star\psi_b\star\psi_c\:\tr(\lambda_a\lambda_b\lambda_c)\right) \nn \\
 & = & \frac{1}{G_s}\left(\frac12 \mbox{ }_{12}\!\bra \hat{V}_2|
|\psi_a\ket_1 Q|\psi_b\ket_2
\:\tr(\lambda_a\lambda_b)
+\frac13 \mbox{ }_{123}\!\bra \hat{V}_3||\psi_a\ket_1 |\psi_b\ket_2 |\psi_c\ket_3
\:\tr(\lambda_a\lambda_b\lambda_c)\right),  
\label{SFTactionCPB}
\eea
where the product $\star$ and the overlap vertices $|\hat{V}_2\ket$ and $|\hat{V}_3\ket$ 
are identical with those constructed in the $U(1)$ case. 
It is clear that, by using the same argument as in the $U(1)$ case, 
we can absorb the Moyal type noncommutativity and obtain the string field theory 
leading to the commutative Yang-Mills theory in the low energy limit. 

 At first sight, it seems that in the form of the string field action (\ref{SFTactionCPB}) 
the noncommutativity originating from the $B$-field background (the product $\star$) 
does not have any connection with that from the nonabelian gauge group (the Chan-Paton 
factor). 
On the other hand, it is known that these two noncommutative natures are related to 
each other in the interesting way: ``Infinitely many coincident D$p$-branes can be 
described by a single D($p+2$)-brane in the background of some constant $B$-field." 
This phenomenon has been discussed in the world sheet formulation of open string theory 
in ref. \cite{Ishibashi1,Kato-Kuroki}. 
So in the framework of Witten's open string field theory constructed here, it might be 
interesting to examine how the noncommutativity from the nonabelian gauge group 
transmutes into that from the constant $B$-field background. 
Because in Matrix theory \cite{BFSS,Banks-Seiberg-Shenker} 
higher dimensional D-branes are constructed from 
the constituent D-particles by utilizing the above fact\footnote{A similar phenomenon 
as the above is known also in the context of 
the IIB matrix model \cite{IKKT}. 
Noncommutative Yang-Mills theory in four dimensions 
arises in expanding matrices around a classical solution 
which represents a condensation of D-instantons 
and composes a D3-brane \cite{AIIKKT}.}, 
it might be useful for clarifying the connection 
between Matrix theory and open string 
field theory, and further it might give a hint for considering 
the covariant and nonperturbative 
formulation of M-theory.

\vspace{3cm}

%%%%%%%%%%%%%%%%%%%%% Acknowledgements %%%%%%%%%%%%%%%%%%%%%%%%%%%%%%%%%%%%%%%

\begin{large}
{\bf Acknowledgements}
\end{large}
\vspace{7mm}

   The author would like to express his gratitude to 
T. Kuroki, Y. Okawa and T. Yoneya for a lot of variable 
discussions about Witten's open string theory and its operator formulation 
for the first half of the year 1997. 
A part of this work was done during his visit at C. E. A. Saclay. 
The author would like to thank members of C. E. A. Saclay for their warm hospitality, 
especially I. Kostov for his kindness.   
Also the author wishes to thank N. Ishibashi and T. Yoneya for helpful discussions. 
The research of the author is supported by the Japan Society for 
the Promotion of Science under the Postdoctral Research Program.

\newpage

\begin{large}
{\bf Appendix}
\end{large} 
\renewcommand{\theequation}{\Alph{section}.\arabic{equation}}
\appendix

%%%%%%%%%%%%%%%%%%%% Appendix A %%%%%%%%%%%%%%%%%%%%%%%%%%%%%%%%

\section{Explicit Form of Overlap Vertices in the Neumann Boundary Condition}
\setcounter{equation}{0}
 Here we give the explicit form of the overlap vertices $|V_N\ket$, which satisfy 
the continuity conditions (\ref{GJ1condition}) and (\ref{GJ2condition}), 
in the case of $N=1,2,3,4$. 
Although it is a slight generalization of the result in the background of 
the flat space-time by Gross and Jevicki \cite{GJ1,GJ2}, we present it 
in order to make this paper more self-contained. 

\subsection{$|I\ket\equiv |V_1\ket$}
The overlap in the $N=1$ case behaves as an identity element for the $*$-product, 
so we denote it by $|I\ket$. Its explicit form is as follows. 
\bea
 & & |I\ket^X \equiv |V_1\ket^X 
  =
\exp \left[-\frac12G^{ij}\sum_{n=0}^{\infty}(-1)^n a^{\dagger}_{n,i}a^{\dagger}_{n,j}
\right]|0\ket,  
\label{identityB=0}
\\
 & & |I\ket^{gh} \equiv |V_1\ket^{gh} 
 =
\exp \left[\sum_{n=2}^{\infty}(-1)^nc_{-n}b_{-n}-2c_0\sum_{k=1}^{\infty}(-1)^kb_{-2k}
   \right. \nn \\
  & &\hspace{5cm} 
   \left. 
+(c_{-1}-c_1)\sum_{k=1}^{\infty}(-1)^kb_{-2k-1}\right]|\Omega\ket.
\eea
Here, the operators $a_{n,i}$ and $a^{\dagger}_{n,i}$ are given by 
\bea
 & & \alpha_{n,i}=\sqrt{n}a_{n,i}, \hspace{1cm}\alpha_{-n,i}=\sqrt{n}a^{\dagger}_{n,i} 
\hspace{1cm}(n\ge 1), \nn \\
 & & x_i=\frac{i}{2}\sqrt{2\alpha'}(a_{0,i}-a^{\dagger}_{0,i}), \hspace{1cm} 
     p_i=\frac{1}{\sqrt{2\alpha'}}(a_{0,i}+a^{\dagger}_{0,i}), 
\eea
the vacuum $|0\ket$ is annihilated by all the $a_{n,i}$'s ($n\ge 0$). 
Also $|\Omega\ket$ in the ghost sector is the so-called 
$SL(2,{\bf R})$ vacuum satisfying 
\beas
 & & b_n|\Omega\ket=0 \hspace{1cm} (n\ge-1), \\
 & & c_n|\Omega\ket=0 \hspace{1cm} (n\ge2). 
\eeas
Note that we usually use the vacua $|\pm\ket$ satisfying 
\bea
 & & b_n|\pm\ket=c_n|\pm\ket=0 \hspace{1cm}(n\ge 1), \\
 & & |+\ket=c_0|-\ket, \hspace{2cm} |-\ket=b_0|+\ket. 
\eea
The vacua $|+\ket$ and $|-\ket$ have the ghost number $\frac12$ and $-\frac12$ 
respectively. Note that the $SL(2,{\bf R})$ vacuum is related to the $|+\ket$ vacuum 
as $|\Omega\ket=b_{-1}b_0|+\ket$, and that it has the ghost number $-\frac32$ which 
is the correct ghost number the identity overlap should have.

\subsection{$|V_2\ket_{12}$}

We write the 2-string overlap for two strings denoted by 1 and 2. 
\bea
 & & |V_2\ket^X_{12}
 = 
\exp \left[-G^{ij}\sum_{n=0}^{\infty}(-1)^n a^{(1)\dagger}_{n,i}a^{(2)\dagger}_{n,j}
\right] |0\ket_{12},  
\label{V2B=0}\\
 & & |V_2\ket^{gh}_{12}
 = 
\exp \left[\sum_{n=1}^{\infty}(-1)^n(c^{(1)}_{-n}b^{(2)}_{-n}+c^{(2)}_{-n}b^{(1)}_{-n})
\right] |\Omega\ket_{12}, 
\eea
where the superscripts $(1)$ and $(2)$ label the strings, the vacuum in 
the ghost sector $|\Omega\ket_{12}$ is defined by 
\beq
|\Omega\ket_{12}=\frac{1}{\sqrt{2}}(|-,+\ket_{1,2}-|+,-\ket_{1,2}).
\eeq

 When considering the connection to the string perturbation theory, 
it is convenient to take the momentum representation rather than the oscillator 
representation with respect to the zero-modes. 
In order to translate into the momentum representation, 
we need the momentum eigenstate constructed from the vacuum of the zero-mode oscillators, 
which is given by 
\beq
|p\ket^{(0)}=\frac{1}{(\det G)^{1/4}}\exp G^{ij}\left[-\frac14 \cdot 2\alpha'p_ip_j
+\sqrt{2\alpha'}p_ia^{\dagger}_{0,j}-\frac12 a^{\dagger}_{0,i}a^{\dagger}_{0,j}\right]|0\ket^{(0)},
\label{peigenstate}
\eeq
where the states of the zero-mode sector were shown by the states with 
the superscript $(0)$. 
Using eq. (\ref{peigenstate}), we can obtain the overlap in the momentum representation 
with respect to the zero-modes: 
\bea
 ^{(0)}_1\!\bra p^{(1)}|\;^{(0)}_2\!\bra p^{(2)}||V_2\ket^X_{12}
 & = & \sqrt{\det G}(\prod_i\delta(p^{(1)}_i+p^{(2)}_i))\exp
\left[-G^{ij}\sum_{n=1}^{\infty}(-1)^n a^{(1)\dagger}_{n,i}a^{(2)\dagger}_{n,j}
\right] |0\ket'_{12}   \nn \\
 & & \times \exp\left[\frac18 \cdot 2\alpha'
G^{ij}(p^{(1)}_i-p^{(2)}_i)(p^{(1)}_j-p^{(2)}_j)\right] \nn \\
 & & \times\left(\prod_{r=1}^2\mbox{ }^{(0)}_r\!\bra p^{(r)}|0\ket^{(0)}_r\right). 
\eea
Here we omitted the irrelevant constant factor multiplying the whole expression. 
The state $|0\ket'_{1,2}$ represents the vacuum with respect to the nonzero-mode  
oscillators, and the ground state wave function of the zero-mode oscillator in the 
momentum representation is 
\beq
^{(0)}\!\bra p|0\ket^{(0)}=\frac{1}{(\det G)^{1/4}}
\exp\left[-\frac14 \cdot 2\alpha'G^{ij}p_ip_j\right]. 
\eeq

\subsection{$|V_4\ket_{1234}$}

We obtain the solution of the continuity conditions assuming the following form: 
\beq
|V_4\ket^X_{1234} = \exp\left[-\frac12G^{ij}\sum_{r,s=1}^4\sum_{n,m=0}^{\infty}
a^{(r)\dagger}_{n,i}V^{rs}_{nm}a^{(s)\dagger}_{m,j}\right]|0\ket_{1234}, 
\eeq
where the matrices $V^{rs}$ are determined by the single matrix $V$ as 
\bea
V^{11}=V^{22}=V^{33}=V^{44} & = & \frac14(V+\bar{V}), \nn \\
V^{12}=V^{23}=V^{34}=V^{41} & = & \frac12C +\frac{i}{4}(V-\bar{V}), \nn \\ 
V^{13}=V^{24}=V^{31}=V^{42} & = & -\frac14(V+\bar{V}), \nn \\
V^{14}=V^{21}=V^{32}=V^{43} & = & \frac12C -\frac{i}{4}(V-\bar{V}), 
\label{relationV}
\eea
with $C_{nm}=(-1)^n\delta_{nm}$. 
$V$ satisfies the following equations, 
which are obtained from the continuity conditions, 
\bea
 & & (1-X)E(1+V)=(1+X)E(1+V^T)=0, \nn \\
 & & (1-X)E^{-1}(1-V^T)=(1+X)E^{-1}(1-V)=0.
\label{equationsV}
\eea
Here the matrices $E$ and $X$ are given by 
\bea
E_{nm}& = & \left\{\begin{array}{ll}1 & \hspace{1cm}(n=m=0) \\
                        \sqrt{2/n} & \hspace{1cm}(n=m\ge 1) \\
                        0 & \hspace{1cm}(\mbox{otherwises}), \end{array}\right. \\
X_{n0} & = & -i\frac{\sqrt{2}}{\pi n}((-1)^n-1)(-1)^{(n-1)/2} \nn \\
       & = & -X_{0n}        \hspace{5cm}(n\ge 1), 
\label{defX1}\\
X_{nm} & = & \frac{i}{\pi}(-1)^{(n-m-1)/2}(1-(-1)^{n+m})
             \left(\frac{1}{n+m}+\frac{1}{n-m}(-1)^m\right) \nn \\
       & = & -X_{mn}       \hspace{5cm} (n,m\ge1,\; n\neq m).
\label{defX2}
\eea
All the other components of $X$ vanish. $X$ is a hermitian matrix having the properties 
\beq
X^T=\bar{X}=CXC=-X,   \hspace{1cm}   X^2=1. 
\label{propertiesX}
\eeq
We can see from eqs. (\ref{equationsV}) and (\ref{propertiesX}) 
that $V$ is hermitian, $V^T=\bar{V}=CVC$ and $V^2=1$. 

Translating to the momentum representation with respect to the zero-modes, we have 
\bea
 & & \left(\prod_{r=1}^4 \mbox{ }^{(0)}_r\!\bra p^{(r)}|\right)|V_4\ket_{1234} \nn \\
 & = & \sqrt{\det G}\prod_i\delta (\sum_{r=1}^4p^{(r)}_i)\exp G^{ij}\left[
-\frac12\sum_{r,s=1}^4\sum_{n,m=1}^{\infty}a^{(r)\dagger}_{n,i}V'^{rs}_{nm}
a^{(s)\dagger}_{m,j}
-\sqrt{2\alpha'}\sum_{r,s=1}^4\sum_{m=1}^{\infty}p^{(r)}_iV'^{rs}_{0m}a^{(s)\dagger}_{m,j}
\right. \nn \\
 & & \left. -\alpha'\sum_{r,s=1}^4p^{(r)}_iV'^{rs}_{00}p^{(s)}_j
-\frac{\alpha'}{8}(p^{(1)}_i-p^{(2)}_i+p^{(3)}_i-p^{(4)}_i)
(p^{(1)}_j-p^{(2)}_j+p^{(3)}_j-p^{(4)}_j)\right]|0\ket'_{1234} \nn \\
 & & \times \left(\prod_{r=1}^4\mbox{ }^{(0)}_r\!\bra p^{(r)}|0\ket^{(0)}_r\right). 
\eea
Here the matrices $V'^{rs}$ are related to the single matrix $V'$ by the same way as in 
(\ref{relationV}), and $V'$ is determined by $V$ as follows: 
\bea
 & & V'_{nm}=V_{nm}+\frac{V_{n0}V_{0m}}{1-V_{00}}, \nn \\
 & & V'_{0m}=\frac{V_{0m}}{1-V_{00}}, \hspace{1cm} V'_{n0}=\frac{V_{n0}}{1-V_{00}}, \nn \\
 & & V'_{00}=\frac{V_{00}}{1-V_{00}} \hspace{1cm} (n,m=1,2,3,\cdots).
\eea

The explicit form of $V'$ can be obtained by using the Neumann function method as in 
ref. \cite{GJ1}. The result is 
\bea
(V'+\bar{V'})_{nm} & = & -2\frac{\sqrt{nm}}{n+m}(u_nu_m-v_nv_m), \nn \\
(V'-\bar{V'})_{nm} & = & \left\{
\begin{array}{ll}2i\frac{\sqrt{nm}}{n-m}(u_nv_m+v_nu_m)  & \hspace{1cm}(n\neq m) \\
                 0                                       & \hspace{1cm}(n=m),     \end{array}
\right. \nn \\
(V'+\bar{V'})_{0m} & = &-\frac{2}{\sqrt{m}}u_m, \nn \\
(V'-\bar{V'})_{0m} & = & -2i\frac{1}{\sqrt{m}}v_m, \nn \\
(V'+\bar{V'})_{n0} & = & -\frac{2}{\sqrt{n}}u_n, \nn \\
(V'-\bar{V'})_{n0} & = & 2i\frac{1}{\sqrt{n}}v_n, \nn \\
V'_{00} & = & \ln 2-\frac12, 
\eea
where $n$ and $m$ run over positive integers. $u_n$ and $v_n$ are given by 
\bea
 & & u_n=\left( \begin{array}{c} -1/2 \\  n/2 \end{array}\right), 
\hspace{1cm} v_n=0, \hspace{1.5cm}(n=2,4,6,\cdots), 
\nn \\
 & & u_n=0,  \hspace{1cm} v_n=\left( \begin{array}{c} -1/2 \\  (n-1)/2
\end{array}\right), \hspace{1.5cm}(n=1,3,5,\cdots).
\eea

For the ghost sector, the result is 
\bea
|V_4\ket^{gh}_{1234} & = & \left.C^2_+(\frac{\pi}{2})C^2_-(\frac{\pi}{2})\exp\sum_{r,s=1}^4(-1)^{r+s}\right[
\sum_{m=1}^{\infty}b_0^{(r)}V'^{rs}_{0m}\sqrt{m}c^{(s)}_{-m}  \nn \\
 & & \left. 
+\sum_{n,m=1}^{\infty}b^{(r)}_{-n}\frac{1}{\sqrt{n}}V'^{rs}_{nm}\sqrt{m}c^{(s)}_{-m}
\right]|\Omega\ket_{1234}. 
\eea
For the $Z_4$-Fourier transformations of $c^{(r)}$'s: 
\bea
C_{\pm} & = & \frac12(ic^{(1)}_{\pm}-c^{(2)}_{\pm}-ic^{(3)}_{\pm}+c^{(4)}_{\pm}), \nn \\
C^2_{\pm} & = & \frac12(-c^{(1)}_{\pm}+c^{(2)}_{\pm}-c^{(3)}_{\pm}+c^{(4)}_{\pm}), \nn \\
\bar{C}_{\pm} & = & \frac12(-ic^{(1)}_{\pm}-c^{(2)}_{\pm}+ic^{(3)}_{\pm}+c^{(4)}_{\pm}), 
\nn \\
C^4_{\pm} & = & \frac12(c^{(1)}_{\pm}+c^{(2)}_{\pm}+c^{(3)}_{\pm}+c^{(4)}_{\pm}) 
\eea
as well as for the same combinations of $b^{(r)}$'s, 
the vacuum $|\Omega\ket_{1234}$ 
is given by the tensor product of the $|-\ket$ vacuum for $(B^2, C^2)$ and 
the three $|+\ket$ 
vacua for $(B^4, C^4)$, $(B,C)$ and $(\bar{B},\bar{C})$. 
The ghost number of the vacuum is $+1$. Note that the mid-point ghost insertion 
$C^2_+(\frac{\pi}{2})C^2_-(\frac{\pi}{2})$ is needed in order for the overlap vertex 
to carry the correct ghost number $+3$.

\subsection{$|V_3\ket_{123}$}
We find the solution of the form 
\beq
|V_3\ket^X_{123}= \exp\left[-\frac12G^{ij}\sum_{r,s=1}^3\sum_{n,m=0}^{\infty}
a^{(r)\dagger}_{n,i}U^{rs}_{nm}a^{(s)\dagger}_{m,j}\right]|0\ket_{123}, 
\label{V3B=0}
\eeq
where the matrices $U^{rs}$ are determined by the single matrix $U$ as 
\bea
U^{11}=U^{22}=U^{33} & = & \frac13(C+U+\bar{U}), \nn \\
U^{12}=U^{23}=U^{31} & = & \frac16(2C-U-\bar{U})+\frac16i\sqrt{3}(U-\bar{U}), \nn \\
U^{13}=U^{21}=U^{32} & = & \frac16(2C-U-\bar{U})-\frac16i\sqrt{3}(U-\bar{U}). 
\label{relationU}
\eea
$U$ satisfies the following equations, 
which is obtained from the continuity conditions, 
\bea
 & & (1-Y)E(1+U)=(1-Y^T)E(1+U^T)=0, \nn \\
 & & (1+Y^T)E^{-1}(1-U^T)=(1+Y)E^{-1}(1-U)=0, 
\label{equationsU}
\eea
with $Y=-\frac12C+\frac{\sqrt{3}}{2}X$ being hermitian and satisfying $Y^2=1$. From 
eqs. (\ref{equationsU}) and the propertires of $Y$, we can see  
that $U$ is hermitian, $U^T=\bar{U}=CUC$  
and $U^2=1$. 

In the momentum representation with respect to the zero-modes, we have 
\bea
 & & \left(\prod_{r=1}^3\mbox{ }^{(0)}_r\!\bra p^{(r)}|\right)|V_3\ket_{123} \nn \\
 & = & \sqrt{\det G}\prod_i\delta(\sum_{r=1}^3p^{(r)}_i)
\exp G^{ij}\left[-\frac12\sum_{r,s=1}^3\sum_{n,m=1}^{\infty}a^{(r)\dagger}_{n,i}
U'^{rs}_{nm}a^{(s)\dagger}_{m,j} \right. \nn \\
 & & \left. -\sqrt{2\alpha'}\sum_{r,s=1}^3\sum_{m=1}^{\infty}p^{(r)}_i
U'^{rs}_{0m}a^{(s)\dagger}_{m,j}-\frac12 \cdot 2\alpha'\sum_{r,s=1}^3p^{(r)}_i
U'^{rs}_{00}p^{(s)}_j\right]|0\ket'_{123} \nn \\
 & & \times \left(\prod_{r=1}^3\mbox{ }^{(0)}_r\!\bra p^{(0)}|0\ket^{(0)}_r\right).
\eea
The matrices $U'^{rs}$'s are also written by the single matrix $U'$. Their relation 
is in the same form as in (\ref{relationU}). 
Also $U'$ is determined by $U$ as 
\bea
 & & U'_{nm}=U_{nm}+\frac{U_{n0}U_{0m}}{1-U_{00}}, \nn \\
 & & U'_{0m}=\frac{U_{0m}}{1-U_{00}}, 
\hspace{1cm} U'_{n0}=\frac{U_{n0}}{1-U_{00}}, \nn \\
 & & U'_{00}=\frac{U_{00}}{1-U_{00}} \hspace{1cm} (n,m=1,2,3,\cdots).
\eea

The explicit form of $U'$ is presented as follows: 

\noindent
For $n\neq m$, $n,m\ge 1$, 
\bea
(U'+\bar{U'})_{nm} & = & \left\{ \begin{array}{ll} 
(-1)^{n+1}\sqrt{nm}\left(\frac{A_nB_m+B_nA_m}{n+m}+\frac{A_nB_m-B_nA_m}{n-m}\right) & 
(n+m:{\rm even}) \\ 0& (n+m:{\rm odd}), \end{array}\right.\nn \\
(U'-\bar{U'})_{nm} & = & \left\{ \begin{array}{ll}
0 & (n+m:{\rm even}) \\ 
i\sqrt{nm}\left(\frac{A_nB_m-B_nA_m}{n+m}+\frac{A_nB_m+B_nA_m}{n-m}\right) 
& (n+m:{\rm odd}), \end{array}\right. \nn \\
(U'+\bar{U'})_{0m} & = & \left\{ \begin{array}{ll} 
-\frac{2}{\sqrt{m}}A_m & \hspace{1cm}(m:{\rm even}) \\ 0 & \hspace{1cm}(m:{\rm odd}),\end{array}
\right. \nn \\
(U'-\bar{U'})_{0m} & = & \left\{ \begin{array}{ll} 
0 & \hspace{1cm}(m:{\rm even}) \\ -i\frac{2}{\sqrt{m}}A_m & \hspace{1cm}(m:{\rm odd}),\end{array}
\right. \nn \\
(U'+\bar{U'})_{n0} & = & \left\{ \begin{array}{ll} 
-\frac{2}{\sqrt{n}}A_n & \hspace{1cm}(n:{\rm even}) \\ 0 & \hspace{1cm}(n:{\rm odd}),\end{array}
\right. \nn \\
(U'-\bar{U'})_{n0} & = & \left\{ \begin{array}{ll} 
0 & \hspace{1cm}(n:{\rm even}) \\ i\frac{2}{\sqrt{n}}A_n & 
\hspace{1cm}(n:{\rm odd}).\end{array}
\right. 
\eea
For the diagonal elements, 
\bea
U_{00} & = & \frac12\ln(3^3/2^4)-\frac12, \nn \\
(U'+\bar{U'})_{nn} & = & (-1)^{n+1}(A_nB_n+1)-n\Delta_n, \nn \\
(U'-\bar{U'})_{nn} & = & 0 \hspace{1cm} (n\ge 1). 
\eea
Here $A_n$ and $B_n$ are given by 
\bea
A_n & = & \left\{\begin{array}{ll} (-1)^{n/2}a_n & \hspace{1cm}(n:{\rm even}) \\
           (-1)^{(n-1)/2}a_n & \hspace{1cm}(n:{\rm odd}),\end{array}\right. \\
B_n & = & \left\{\begin{array}{ll} (-1)^{n/2}b_n & \hspace{1cm}(n:{\rm even}) \\
           (-1)^{(n-1)/2}b_n & \hspace{1cm}(n:{\rm odd}),\end{array}\right. 
\eea
\beq
\left(\frac{1+x}{1-x}\right)^{1/3}=\sum_{n=0}^{\infty}a_nx^n, \hspace{1cm}
\left(\frac{1+x}{1-x}\right)^{2/3}=\sum_{n=0}^{\infty}b_nx^n. 
\eeq
Also $\Delta_n$'s are 
\beq
\Delta_n=\left\{\begin{array}{ll}
-\frac{\sqrt{3}}{\pi}(-1)^{n/2}(\tilde{O}^a_nB_n-\tilde{O}^b_nA_n) 
& (n=2,4,6,\cdots) \\
\frac{\sqrt{3}}{\pi}(-1)^{(n-1)/2}(\tilde{E}^a_nB_n-\tilde{E}^b_nA_n) 
& (n=1,3,5,\cdots) ,
\end{array}\right. 
\eeq
where $\tilde{O}^a_n$, $\tilde{E}^a_n$, $\tilde{O}^b_n$ and $\tilde{E}^b_n$ can be 
written in terms of $a_n$'s and $b_n$'s as 
\beas
& & \tilde{O}^a_0=\frac{\pi\sqrt{3}}{4}\ln\frac{3^3}{2^4}, \hspace{1cm}
    \tilde{E}^a_1=\frac{\pi}{\sqrt{3}}\left(\ln\frac32+\frac16\right), \\
& & \tilde{E}^a_n=\frac{\pi\sqrt{3}}{2}\left[\left(\ln\frac32+\frac16\right)a_n-a_{n+1}
\right.\\
& & \hspace{2cm}\left.+\sum_{l=0}^{n-1}\frac{(-1)^l}{l+1}a_{n-l-1}(a_{l+2}+\frac12a_l)
\right]\hspace{1cm}(n=3,5,7,\cdots), \\
& & \tilde{O}^a_n=\frac{\pi\sqrt{3}}{2}\left[\left(\ln\frac32+\frac16\right)a_n-a_{n+1}
\right.\\
& & \hspace{2cm}\left.+\sum_{l=0}^{n-1}\frac{(-1)^l}{l+1}a_{n-l-1}(a_{l+2}+\frac12a_l)
\right]\hspace{1cm}(n=2,4,6,\cdots), \\
& & \tilde{O}^b_0=\frac{\pi\sqrt{3}}{4}\ln 3, \hspace{1cm}
    \tilde{E}^b_1=\frac{2\pi}{\sqrt{3}}\left(\ln\frac32-\frac{1}{12}\right), \\
& & \tilde{E}^b_n = \frac{\pi\sqrt{3}}{4}\left[\left(2\ln\frac32-\frac16\right)b_n-b_{n+1}
\right.\\
& & \hspace{2cm}\left.+\sum_{l=0}^{n-1}\frac{(-1)^l}{l+1}b_{n-l-1}(a_{l+2}-\frac12b_l)
\right]\hspace{1cm}(n=3,5,7,\cdots), \\
& & \tilde{O}^b_n= \frac{\pi\sqrt{3}}{4}\left[\left(2\ln\frac32-\frac16\right)b_n-b_{n+1}
\right.\\
& & \hspace{2cm}\left.+\sum_{l=0}^{n-1}\frac{(-1)^l}{l+1}b_{n-l-1}(a_{l+2}-\frac12b_l)
\right]\hspace{1cm}(n=2,4,6,\cdots). 
\eeas

Finally we show the result of the ghost 3-string overlap vertex: 
\beq
|V_3\ket^{gh}_{123} = \exp\sum_{r,s=1}^3\left[\sum_{n,m=1}^{\infty}
b^{(r)}_{-n}\frac{1}{\sqrt{n}}W^{rs}_{nm}\sqrt{m}c^{(s)}_{-m} 
+\sum_{m=1}^{\infty}b^{(r)}_0W^{rs}_{0m}
\sqrt{\frac{m}{2}}c^{(s)}_{-m}\right]|\Omega\ket_{123}, 
\eeq
where the matrices $W^{rs}$'s are related to the single matrix $W$ as 
\bea
W^{11}=W^{22}=W^{33} & = & \frac13(-C+W+\bar{W}), \nn \\
W^{12}=W^{23}=W^{31} & = & \frac16(-2C-W-\bar{W})+i\frac{\sqrt{3}}{6}(W-\bar{W}), \nn \\
W^{13}=W^{21}=W^{32} & = & \frac16(-2C-W-\bar{W})-i\frac{\sqrt{3}}{6}(W-\bar{W}),
\eea
which is different from the relation for $U$ (\ref{relationU}) by the sign 
in front of $C$. 
The explicit from of $W$ is given as follows: 

\noindent
For $n,m \ge 1$ and $n\neq m$, 
\bea
 & &(W+\bar{W})_{nm} = \left\{ \begin{array}{ll} 
(-1)^{n+1}\sqrt{nm}\left(\frac{A_nB_m+B_nA_m}{n+m}-\frac{A_nB_m-B_nA_m}{n-m}\right) & 
(n+m:{\rm even}) \\ 0& (n+m:{\rm odd}), \end{array}\right.\nn \\
 & & (W-\bar{W})_{nm} = \left\{ \begin{array}{ll}
0 & (n+m:{\rm even}) \\ 
i\sqrt{nm}\left(\frac{A_nB_m-B_nA_m}{n+m}-\frac{A_nB_m+B_nA_m}{n-m}\right) 
& (n+m:{\rm odd}), \end{array}\right. \nn \\
 & & (W+\bar{W})_{0m} =\left\{ \begin{array}{ll} 
-2\sqrt{\frac{2}{m}}B_m & \hspace{1cm}(m:{\rm even}) \\ 0 & \hspace{1cm}(m:{\rm odd}),\end{array}
\right. \nn \\
 & & (W-\bar{W})_{0m} =  \left\{ \begin{array}{ll} 
0 & \hspace{1cm}(m:{\rm even}) \\ i2\sqrt{\frac{2}{m}}B_m & \hspace{1cm}(m:{\rm odd}),\end{array}
\right. \nn \\
 & & W_{n0}=\bar{W}_{n0}=0. 
\eea
For the diagonal elements, 
\bea
& & W_{00}=\bar{W}_{00}=0, \nn \\
& & (W+\bar{W})_{nn}=(-1)^n(1-a_nb_n)+n\Delta_n, \nn \\
& & (W-\bar{W})_{nn}=0 \hspace{1cm}(n\ge 1). 
\eea
The vacuum $|\Omega\ket_{123}$ is the tensor product of the three vacua $|+\ket_r$ 
($r=1,2,3$), whose ghost number is $+\frac32$. This is the correct ghost number which 
the three-string overlap vertex should have.

\newpage
 
%%%%%%%%%%%%%%%%%%%%%% References %%%%%%%%%%%%%%%%%%%%%%%%%%%%%%%%

\end{document}